\documentclass[12pt, draftclsnofoot, onecolumn]{IEEEtran}

\IEEEoverridecommandlockouts

\ifCLASSOPTIONcompsoc
\usepackage[caption=false,font=normalsize,labelfon
t=sf,textfont=sf]{subfig}
\else
\usepackage[caption=false,font=footnotesize]{subfi
g}
\fi

\usepackage{amsmath,amssymb}
\usepackage{multicol}
\usepackage{graphicx,graphics,color,psfrag}
\usepackage{cite,balance}
\usepackage{algorithm}
\usepackage{accents}

\usepackage{bm}
\usepackage{url}
\usepackage{algorithmic}
\usepackage[english]{babel}
\usepackage{multirow}
\usepackage{enumerate}
\usepackage{cases}
\usepackage{stfloats}
\usepackage{dsfont}
\usepackage{color,soul}
\usepackage{amsfonts}
\usepackage{tcolorbox}
\usepackage{amsmath}
\usepackage{float}

\usepackage{cite,graphicx,amsmath,amssymb}
\usepackage{fancyhdr}
\usepackage{hhline}
\usepackage{graphicx,graphics}
\usepackage{array,color}
\usepackage{amsmath}
\usepackage{stfloats}
\usepackage[flushleft]{threeparttable}
\usepackage{booktabs}

\newtheorem{proposition}{Proposition}

\ifCLASSINFOpdf

\else

\fi

\begin{document}
\title{Cram\'er-Rao Bound Minimization for IRS-Enabled Multiuser Integrated Sensing and Communications}

\author{Xianxin~Song,  Xiaoqi~Qin, Jie~Xu, and Rui~Zhang
\thanks{Part of this paper was presented at the IEEE International Conference on Communications (ICC) 2023 \cite{xianxin2023ICC}.
}
\thanks{X. Song and J. Xu are with the School of Science and Engineering (SSE) and the Future Network of Intelligence Institute (FNii),  The Chinese University of Hong Kong (Shenzhen), Shenzhen 518172, China (e-mail: xianxinsong@link.cuhk.edu.cn, xujie@cuhk.edu.cn). J. Xu is the corresponding author.}
\thanks{X. Qin is with the State Key Laboratory of Networking and Switching Technology, Beijing University of Posts and Telecommunications, Beijing 100876, China (e-mail: xiaoqiqin@bupt.edu.cn).}
\thanks{R. Zhang is with School of Science and Engineering, Shenzhen Research Institute of Big Data, The Chinese University of Hong Kong, Shenzhen, Guangdong 518172, China (e-mail: rzhang@cuhk.edu.cn). He is also with the Department of Electrical and Computer Engineering, National University of Singapore, Singapore 117583 (e-mail: elezhang@nus.edu.sg).}
}

\maketitle
\begin{abstract}
This paper investigates an intelligent reflecting surface (IRS) enabled multiuser integrated sensing and communications (ISAC) system, which consists of one multi-antenna base station (BS), one IRS, multiple single-antenna communication users (CUs), and one target at the non-line-of-sight (NLoS) region of the BS. The IRS is deployed to not only assist the communication from the BS to the CUs, but also enable the BS's NLoS target sensing  based on the echo signals from the BS-IRS-target-IRS-BS link. We consider two types of targets, namely the extended and point targets, for which the BS aims to estimate the complete target response matrix and the target direction-of-arrival (DoA) with respect to the IRS, respectively. To provide full degrees of freedom for sensing, we consider that the BS sends dedicated sensing signals in addition to the communication signals. Accordingly, we model two types of CU receivers, namely Type-I and Type-II CU receivers, which do not have and have the capability of canceling the interference from the sensing signals, respectively. Under each setup, we jointly optimize the transmit beamforming at the BS and the reflective beamforming at the IRS to minimize the Cram\'er-Rao bound (CRB) for target estimation, subject to the minimum signal-to-interference-plus-noise ratio (SINR) constraints at the CUs and the maximum transmit power constraint at the BS. We present efficient algorithms to solve the highly non-convex SINR-constrained CRB minimization problems, by using the techniques of alternating optimization, semi-definite relaxation, and successive convex approximation. Numerical results show that the proposed design achieves lower estimation CRB than other benchmark schemes, and the sensing signal interference cancellation at Type-II CU receivers is beneficial when the number of CUs is greater than one.
\end{abstract}


\IEEEpeerreviewmaketitle

\section{Introduction}
Integrated sensing and communications (ISAC) has been recognized as one of the key technologies for the future sixth-generation (6G) wireless networks, in which base station (BS) infrastructures and spectrum resources are used to provide ubiquitous sensing and communication services at the same time (see, e.g., \cite{liu2021integrated,9705498} and the references therein). In practice, the performance of ISAC networks is highly dependent on the wireless propagation environment. For instance, the wireless sensing generally relies on the line-of-sight (LoS) links for realizing the estimation of target angles and ranges \cite{richards2014fundamentals}, and the wireless communications demand strong communication channels and weak interference for meeting the quality-of-service (QoS) or signal-to-interference-plus-noise ratio (SINR) requirements. However, due to the obstacles such as buildings/trees in outdoor scenarios and furniture/walls in indoor scenarios, the signal propagation links between BSs and sensing targets/communication users (CUs) may be significantly attenuated or even blocked, thus severely degrading the sensing and/or communication performance.

Recently, intelligent reflecting surface (IRS)\cite{8811733,9326394}, also known as reconfigurable intelligent surface (RIS)\cite{9424177}, has emerged as a promising technology to resolve the above issues by reconfiguring the wireless propagation environment via properly adjusting the phase shifts of incident signals by controlling the reflecting elements. On one hand, IRSs can improve the communication performance by, e.g., enhancing the received signal strength at CUs, refining channel ranks by providing more signal paths, and suppressing the co-channel interference\cite{9326394}. On the other hand, IRSs can also enhance the wireless sensing performance via creating new virtual LoS links for the targets located in the blocked region \cite{xianxin} and providing additional sensing angles for the targets in the LoS region\cite{Stefano}. Therefore, IRS-enabled ISAC has received growing  interests recently \cite{chepuri2022integrated,10077119,song2021joint,sankar2022beamforming,9724202,9769997,9729741,9591331,10086570,wang2022stars}.

Despite its appealing benefits, IRS-enabled ISAC brings new challenges to be dealt with. First, as compared to the conventional ISAC system without IRSs, the reflective beamforming at IRSs (via controlling the reflection phase shifts) introduces a new design degrees of freedom (DoFs) for enhancing the ISAC performance. However, the design of reflective beamforming is challenging due to the large number of reflection phase shifts that need to be jointly optimized and their resultant unit-modulus constraints on the reflection coefficients \cite{8811733}. Next, as sensing and communication tasks coexist in the ISAC system, it is important to properly manage their co-channel interference and balance their performance tradeoff via the joint transmit and reflective beamforming design. This, however, is also challenging, as the resultant optimization problems are highly non-convex and involve  complicated coupling between the transmit and reflective beamformers. 

In the literature, prior works have investigated IRS-enabled ISAC under different setups (see, e.g., \cite{chepuri2022integrated,10077119,song2021joint,sankar2022beamforming,9724202,9769997,9729741,9591331,10086570,wang2022stars}). For example, the authors in \cite{song2021joint} considered an IRS-enabled ISAC system with multiple targets at the non-LoS (NLoS) region of the BS and one CU, in which the minimum sensing beampattern gain (or worst-case target illumination power) for target sensing was maximized while ensuring the signal-to-noise ratio (SNR) requirement at the CU. The authors in \cite{sankar2022beamforming} studied a double-IRS-enabled ISAC system with multiple CUs, where one IRS was deployed to help the CUs' communications and the other IRS was equipped to assist the target sensing. The authors studied two problems of minimizing the sensing cost function and maximizing the worst-case target illumination power, respectively, while ensuring the minimum SINR requirements at individual CUs. The authors in \cite{9724202} deployed dedicated sensing receivers at the IRS and specialized transmitters near the IRS, such that the target direction-of-arrival (DoA) estimation is performed at the IRS receivers based on echo signals from the transmitter-target-IRS receivers and the transmitter-IRS-target-IRS receivers link. The authors maximized the average received signals power through the specialized transmitters-IRS-target-IRS receivers and transmitter-target-IRS receivers links at the IRS by reflective beamforming design. The authors in \cite{9769997} considered the IRS-enabled ISAC system by utilizing both the direct BS-target-BS and the reflected BS-IRS-target-IRS-BS links for target sensing, in which the joint beamforming was optimized to maximize the sensing SINR subject to the QoS requirements for communications, by considering different design principles based on zero-forcing (ZF), minimum mean square error (MMSE), and constructive interference (CI), respectively. The authors in \cite{9729741} considered a double-IRS-assisted communication-radar coexistence system, in which one IRS is placed close to the BS transmitter and the other is near the CU receiver for assisting communication only. Under this setup, the communication SINR was maximized while guaranteeing the sensing SINR. 

%Furthermore, the authors in \cite{yu2023active} considered an active IRS-aided ISAC system to overcome the double-fading problem of the reflected link, in which the sensing SINR was maximized under the minimum SINR requirement at CU.

%In the literature, there have been several prior works investigating IRS-enabled ISAC \cite{chepuri2022integrated,10077119,sankar2022beamforming,9729741,song2021joint,9769997,yu2023active,9591331,10086570,wang2022stars} under different setups. For instance, in \cite{sankar2022beamforming,9729741}, one or more IRS were used to only assist the communication or sensing functionality. In \cite{song2021joint,9769997}, one IRS was deployed to assist target sensing and communication simultaneously, by utilizing only the reflected BS-IRS-target-IRS-BS link or both the direct BS-target-BS and the reflected BS-IRS-target-IRS-BS links for sensing, respectively. Furthermore, the authors in \cite{yu2023active} considered an active IRS-aided ISAC system to overcome the double-fading problem of the reflected link.

Besides using the target illumination power and the sensing SINR as the sensing performance metrics in \cite{sankar2022beamforming,9729741,song2021joint,9769997,9724202}, there has been another line of works on IRS-enabled ISAC that adopted the Cram\'er-Rao bound (CRB)  as the sensing performance metric for estimating target parameters (e.g., target locations and angles)\cite{9591331,10086570,wang2022stars}. The CRB provides a lower bound on the variance of any unbiased parameter estimators, which serves as the fundamental performance limits for parameter estimation \cite{kay1993fundamentals,bekkerman2006target,9652071,huaCRB-RATE}. For example, prior works \cite{9591331,10086570} considered the IRS-enabled ISAC systems with IRSs deployed for assisting communications only, in which the sensing is performed at the BS based on the BS-target-BS links. In this case, the authors optimized the communication performance (in terms of multi-user interference minimization or sum rate maximization), subject to the maximum CRB constraint for target DoA estimation at the BS. The authors in \cite{wang2022stars} considered another type of semi-passive IRS sensing, in which dedicated sensing receivers were deployed at the IRS, such that the sensing is performed at IRS based on the BS-IRS-target-IRS links. Under this setup, the authors minimized the CRB for target DoA estimation at the IRS, subject to the minimum SINR requirements at multiple CUs.
%However, prior works \cite{9591331,10086570} used IRS to assist communication only and performed sensing based on the direct BS-target-BS links, and \cite{wang2022stars} performed sensing at IRS based on the BS-IRS-target-IRS links via deploying additional receivers at the IRS. 
%However, these prior works are not applicable to the practical scenario when the IRS is purely passive (without dedicated sensing receivers) such that  the NLoS target sensing needs to be performed at the BS through the BS-IRS-target-IRS-BS link\cite{xianxin}. This thus motivates our work in this paper. 

Different from the above prior works that considered IRS-enabled ISAC with BS LoS target sensing \cite{9591331,10086570} and semi-passive IRS sensing \cite{wang2022stars}, in this paper we focus on the CRB minimization in an IRS-enabled ISAC system with passive IRS sensing (i.e., the IRS is not equipped with dedicated sensing receivers for cost consideration), such that the NLoS target sensing is performed at the BS based on the BS-IRS-target-IRS-BS link \cite{xianxin}. More specifically, we consider an IRS-enabled multiuser ISAC system with one BS, one IRS, multiple CUs, and a target at the NLoS region of the BS. The IRS is deployed to not only assist the wireless communication from the BS to the CUs, but also create a virtual LoS channel over the BS-IRS-target-IRS-BS link to facilitate the target sensing at the BS. First, we consider two types of targets according to the spatial extent of target, namely the extended and point targets, respectively\cite{xianxin,259642,9652071,huaCRB-RATE,4200705,8579200}. For the extended target, the BS estimates the complete target response matrix with respect to (w.r.t.) the IRS (or equivalently the cascaded IRS-target-IRS channel matrix) as unknown parameters. In contrast, for the point target, the BS estimates the target DoA and channel coefficient w.r.t. the IRS as unknown parameters. Moreover, in order to achieve full DoFs for target sensing, we consider that the BS transmits dedicated sensing signals in addition to the communication signals \cite{9124713,9652071,hua}. The dedicated sensing signals are pseudorandom or deterministic sequences, which are unknown/known to the CUs, respectively. As a result, we model two types of CU receivers, namely Type-I and Type-II CU receivers, which do not have and have the capability of canceling the interference from the sensing signals, respectively.  

Under the above two target models and two types of CU receivers, we aim to minimize the CRB for estimating the parameters of interest (i.e., the target response matrix in the extended target model and the target DoA in the point target model), by jointly optimizing the transmit beamforming at the BS and the reflective beamforming at the IRS, subject to the minimum SINR requirements at individual CUs.
The formulated SINR-constrained CRB minimization problems are highly non-convex due to the non-convex CRB objective functions, the unit-modulus constraints on reflective beamforming,  and the coupled relation between the transmit and reflective beamforming. To solve such non-convex SINR-constrained CRB minimization problems, we present efficient algorithms to obtain converged solutions via alternating optimization, semi-definite relaxation (SDR), and successive convex approximation (SCA). 
%For the extended target case, the CRB for target response matrix estimation only depends on the transmit beamformers at the BS. In order to solve the non-convex problems efficiently, we decompose them into two subproblems, i.e., the transmit beamforming design and the reflective beamforming design problems, and iteratively solving them in an alternating manner. The transmit beamforming design problem is optimally solved by using the SDR technique. The reflected beamforming design problem is a feasibility-check problem, which is solved by using the SDR and Gaussian randomization techniques.
% To deal 
%  The formulated SINR-constrained CRB minimization problem is non-convex and difficult to solve. 
%We present an efficient algorithm via alternating optimization, semi-definite relaxation (SDR), and successive convex approximation (SCA). For the extended target case, the resultant CRB minimization problem is convex, in which only the transmit beamforming vectors at the AP are optimization variables.
Finally, numerical results show that the proposed design achieves improved sensing performance in terms of minimized CRB, as compared to other benchmark schemes with transmit beamforming only and random IRS phase shifts, and a separate communication and sensing design. Furthermore, it is shown that if there is only one CU, then Type-II CU receiver achieves the same performance as Type-I CU receiver, as the dedicated sensing signals are not needed in this case. In contrast, if there are more than one CUs, then Type-II CU receiver leads to significant performance gains over Type-I CU receiver, thus validating the benefit of sensing signal interference cancellation in this case.

%It is also shown that the case with Type-II CU receivers leads to significant performance gains over that with Type-I CU receivers when the number of CUs is greater than one, thus validating the benefit of sensing signal interference cancellation. 

%Furthermore, with an extended target, the BS aims to estimate the complete target response matrix with respect to (w.r.t.) the IRS based on the echo signals from the BS-IRS-target-IRS-BS link, for which the estimation CRB has been derived in \cite{xianxin}. Building upon this, we jointly optimize the transmit beamforming at the BS and the reflective beamforming at the IRS to minimize the estimation CRB, subject to the minimum SINR constraints at the CUs and the maximum transmit power constraint at the BS. Although the formulated SINR-constrained CRB minimization problems are highly non-convex, we present efficient algorithms to obtain converged solutions by using alternating optimization and semi-definite relaxation (SDR). Numerical results show that the proposed design achieves improved sensing performance in terms of lower CRB, as compared to other benchmark schemes. It is also shown that the case with Type-II CU receivers leads to significant performance gains over that with Type-I CU receivers when the number of CUs is greater than one, thus validating the benefit of sensing signal interference pre-cancellation.

\textit{Notations:} 
Boldface letters refer to vectors (lower case) or matrices (upper case). For a square matrix $\mathbf S$, $\mathrm {tr}(\mathbf S)$ and $\mathbf S^{-1}$ denote its trace and inverse, respectively, and $\mathbf S \succeq \mathbf{0}$ means that $\mathbf S$ is positive semi-definite. For an arbitrary-size matrix $\mathbf M$, $\mathrm {rank}(\mathbf M)$, $\mathbf M^*$, $\mathbf M^{T}$, and $\mathbf M^{H}$ denote its rank, conjugate, transpose, and conjugate transpose, respectively. We use $\mathcal{C N}(\mathbf{0}, \mathbf{\Sigma})$ to denote the distribution of a circularly symmetric complex Gaussian (CSCG) random vector with mean vector $\mathbf 0$ and covariance matrix $\mathbf \Sigma$, and $\sim$ to denote “distributed as”. The spaces of $x \times y$ complex and real matrices are denoted by $\mathbb{C}^{x \times y}$ and  $\mathbb{R}^{x \times y}$, respectively. The real and imaginary parts of a complex number are denoted by $\mathrm{Re}\{\cdot\}$ and $\mathrm{Im}\{\cdot\}$, respectively.
The symbol $\mathbb{E}(\cdot)$ denotes the statistical expectation, $\|\cdot\|$ denotes the Euclidean norm, $|\cdot|$ denotes the magnitude of a complex number, $\mathrm {diag}(a_1,\cdots,a_N)$ denotes a diagonal matrix with diagonal elements $a_1,\cdots,a_N$, $\otimes$ denotes the Kronecker product,  $\mathrm{vec}(\cdot)$ denotes the vectorization operator, and $\mathrm {arg}(\mathbf x)$ denotes a vector with each element being the phase of the corresponding element in $\mathbf x$.

%The space of $x \times y$ complex matrices is denoted by $\mathbb{C}^{x \times y}$. The symbol $\|\cdot\|$ stands for the Euclidean norm, $|\cdot|$ for the magnitude of a complex number, and $\mathrm {diag}(a_1,\cdots,a_N)$ for a diagonal matrix with diagonal elements $a_1,\cdots,a_N$.

% \begin{figure}[htbp]
% 	\centering
% 	\begin{minipage}{0.43\linewidth}
% 		\centering
% 		\includegraphics[width=0.9\linewidth]{Fig/CRB_vs_SNR_K_1.eps}
% 		 \caption{The estimation CRB versus the SINR threshold $\Gamma$ with $K=1$ in Scenario I.}
%     	\label{fig:CRB_SNR_single}
% 	\end{minipage}
% 	\quad
% 	\begin{minipage}{0.43\linewidth}
% 		\centering
% 		\includegraphics[width=0.9\linewidth]{Fig/CRB_vs_SNR_K_3.eps}
% 		\caption{The estimation CRB versus the SINR threshold $\Gamma$ with $K=3$  in Scenario I.}
%     	\label{fig:CRB_SNR_multi}
% 	\end{minipage}
% \end{figure}

% \begin{figure}[t]
%     \centering
%     \includegraphics[width=0.4\textwidth]{Fig/system_model.pdf}
%     \caption{System model of IRS-enabled multiuser ISAC.}
%     \label{system_model}
% \end{figure}
\section{System Model}
\begin{figure}[t]
\centering
\subfloat[Extended target case]{\includegraphics[width=2.5in]{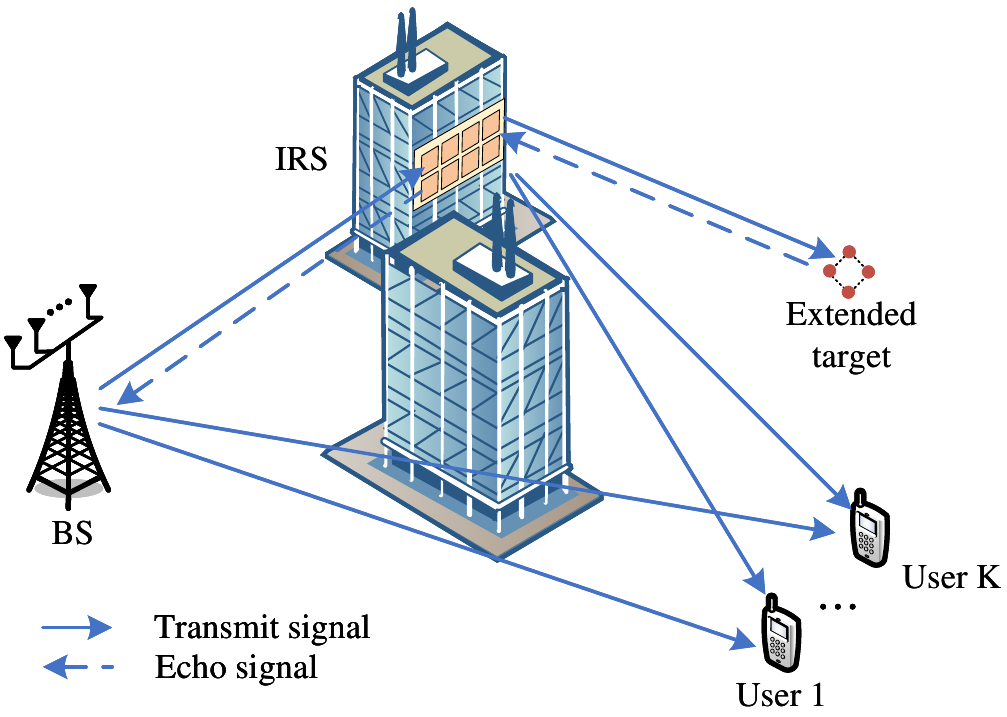}%
\label{Extended target}}\qquad
\subfloat[Point target case]{\includegraphics[width=2.5in]{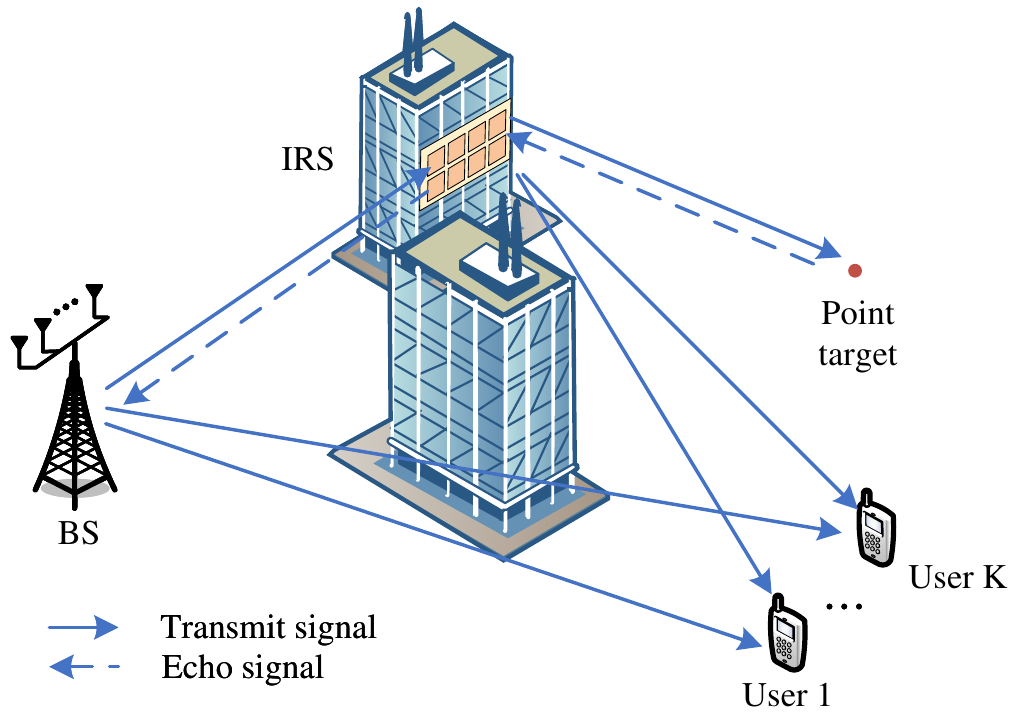}%
\label{Point target}}
\caption{System model of IRS-enabled multiuser ISAC.}
\label{system_model}
\end{figure}

We consider an IRS-enabled ISAC system as shown in Fig.~\ref{system_model}, which consists of one BS with $M>1$ antennas, one IRS with $N>1$ uniform linear array (ULA) reflecting elements, $K \ge 1$ single-antenna CUs, and one sensing target (i.e., extended target in Fig.~\ref{Extended target} and point target in Fig.~\ref{Point target}) at the NLoS region of the BS (i.e., the LoS path between the target and the BS is severely blocked). Let $\mathcal{K}=\{1,\cdots,K\}$ denote the set of CUs, and $\mathcal{N}=\{1,\cdots,N\}$ denote the set of reflecting elements at the IRS.

We consider one particular ISAC transmission block consisting of $T$ symbols. Let $\mathcal{T} = \{1, \cdots, T\}$ denote the set of symbols. To achieve full DoFs for sensing, we assume that the BS sends both communication/information signals and dedicated sensing signals for ISAC\cite{9124713,9652071,hua}. Let $s_k(t)$ denote the transmit communication signal for CU $k$ at symbol $t$, and $\mathbf w_k$ denote the corresponding transmit beamforming vector. Here, $s_k(t)$'s are assumed to be independent and identically distributed (i.i.d.) random variables with zero mean and unit variance. Let $\mathbf x_0(t) \in\mathbb{C}^{M \times 1}$ denote the dedicated sensing signal vector at symbol $t$, which is generated independently from the communication signals $s_k(t)$'s. The sample covariance matrix of $\mathbf x_0(t)$ is 
\begin{equation}
\mathbf R_0 \triangleq \frac{1}{T}\sum_{t\in\mathcal{T}}\mathbf{x}_0(t)\mathbf{x}_0^{H}(t) \succeq \mathbf{0}.
\end{equation}
Suppose that $\mathrm{rank}(\mathbf R_0) = l_0, l_0\leq \min(T, M) $ and the eigenvalue decomposition (EVD) of $\mathbf R_0$ is given by $\mathbf R_0=\mathbf U \mathbf \Lambda \mathbf U^{H}$, 
where $\mathbf \Lambda = \mathrm{diag}(\lambda_1, \cdots, \lambda_M)$ and $\mathbf U = [\mathbf u_1,\cdots ,\mathbf u_M]$ with $\mathbf U  \mathbf U^{H} = \mathbf U^{H} \mathbf U= \mathbf{I}_{M}$. Accordingly, we have $\lambda_1 \ge \cdots \ge \lambda_{l_0} > \lambda_{l_0+1}= \cdots =\lambda_{M}= 0$. This means that there are a number of $l_0$ sensing beams transmitted by the BS, each of which is denoted by $\sqrt{\lambda_i}\mathbf u_i, i\in\{1, \cdots, l_0\}$.
Then the transmitted signal by the BS at symbol $t \in \mathcal{T}$ is expressed as 
\begin{equation}
\mathbf x(t) =\sum_{k\in\mathcal{K}}\mathbf w_k s_k(t) + \mathbf{x}_0(t).
\end{equation}
By assuming $T$ to be sufficiently large, we consider that the sample covariance matrix of the transmitted signal $\mathbf x(t)$ is same as its statistical covariance matrix, i.e., 
\begin{equation}\label{eq:approx_R}
\mathbf R_x \triangleq \frac{1}{T}\hspace{-0.2em}\sum_{t\in\mathcal{T}}\mathbf{x}(t)\mathbf{x}^{H}(t)
 \approx \mathbb{E}(\mathbf{x}(t)\mathbf{x}^{H}(t))= \sum_{k\in\mathcal{K}} \mathbf w_k\mathbf w_k^{H} + \mathbf R_0.
\end{equation}
Let $P_0$ denote the maximum transmit power at the BS. We thus have 
\begin{equation}\label{equ:sum_power_constr}
  \mathbb{E}(\| \mathbf x(t) \|^2) =\sum_{k\in\mathcal{K}} \|\mathbf w_k\|^2 + \mathrm {tr}(\mathbf R_0) \le P_0.
\end{equation}
Furthermore, we consider that the IRS employs the reflective beamforming to facilitate the ISAC operation. In particular, the IRS can adjust the phase shifts at the reflecting elements with unit amplitudes\cite{8811733}. Let $\mathbf v = [e^{j\phi_1},\cdots,e^{j\phi_{N}}]^{T}$ denote the vector collecting the $N$ reflecting coefficients at the IRS, where $\phi_n \in (0, 2\pi]$ being the phase shift of each element. Furthermore, let $\mathbf{\Phi} = \mathrm {diag}(\mathbf v)$ denote the corresponding  reflection matrix.
\subsection{Communication Model}
First, we consider the wireless communication from the BS to the CUs. Let $\mathbf h_{\text{d},k}^{H}\in \mathbb C^{1\times M}$ and $\mathbf h_{\text{r},k}^{H}\in \mathbb C^{1\times N}$ denote the channel vectors from the BS and the IRS to CU $k$, respectively. Let $\mathbf G \in \mathbb C^{N\times M}$ denote the channel matrix from the BS to the IRS. We assume that the BS knows the perfect channel state information (CSI) via proper channel estimation methods\cite{9722893}. This assumption is made in order to characterize the fundamental ISAC performance. The received signal by CU $k \in \mathcal{K}$ at symbol $t \in \mathcal{T}$ is
\begin{equation}
\begin{split}
&y_k(t) =  (\mathbf h_{\text{d},k}^{H}+ \mathbf h_{\text{r},k}^{H} \mathbf \Phi\mathbf G )\mathbf x(t)+  n_k(t)\\
 =& \underbrace{(\mathbf h_{\text{d},k}^{H}+ \mathbf h_{\text{r},k}^{H} \mathbf \Phi\mathbf G )\mathbf w_k s_k(t)}_\text{desired information signal}+ \underbrace{(\mathbf h_{\text{d},k}^{H}+ \mathbf h_{\text{r},k}^{H} \mathbf \Phi\mathbf G )\hspace{-0.6em}\sum_{{i\in\mathcal{K},i\neq k}}\hspace{-0.6em}\mathbf w_i s_i(t)}_\text{inter-user interference} +  \underbrace{(\mathbf h_{\text{d},k}^{H}+ \mathbf h_{\text{r},k}^{H} \mathbf \Phi\mathbf G )\mathbf x_0(t)}_\text{sensing signal interference} + n_k(t),  
 \end{split}
\end{equation}
where $n_k(t) \sim \mathcal{CN}(0,\sigma_k^2)$ denotes the noise at the receiver of CU $k$, which may include the background interference. Note that the dedicated sensing signals $\mathbf x_0(t)$'s are pseudorandom or deterministic sequences that are assumed to be unknown/known to the CUs, respectively. As a result, we consider two types of CU receivers, namely Type-I and Type-II CU receivers, which cannot cancel and can cancel the  pseudorandom/deterministic interference caused by the sensing signals $\mathbf x_0(t)$, respectively \cite{9124713,9652071,hua}. For the cases with Type-I and Type-II CU receivers, the corresponding SINRs at CU $k \in \mathcal K$ are respectively given by 
% \begin{equation}\label{eq:SINR}
% \gamma_k^{(\text{I})}=\frac{|\mathbf h_k^{H} \mathbf w_k|^2}{\sum_{i\in\mathcal{K},i\neq k}|\mathbf h_k^{H}\mathbf w_i|^2+\mathbf h_k^{H}\mathbf R_0 \mathbf h_k +\sigma_k^2}, \forall k \in \mathcal{K},
% \end{equation}
% \begin{equation}\label{eq:SINR_II}
% \gamma_k^{\text{II}}= \frac{|\mathbf h_k^{H}\mathbf w_k|^2}{\sum_{i\in\mathcal{K},i\neq k}|\mathbf h_k^{H}\mathbf w_i|^2+ \sigma_k^2}, \forall k \in \mathcal{K},
% \end{equation}
\begin{equation}\label{eq:SINR}
\gamma_k^{\text{I}}=\frac{|(\mathbf h_{\text{d},k}^H+ \mathbf h_{\text{r},k}^H  \mathbf \Phi \mathbf G) \mathbf w_k|^2}{\sum_{i\in\mathcal{K},i\neq k}|(\mathbf h_{\text{d},k}^H+ \mathbf h_{\text{r},k}^H  \mathbf \Phi \mathbf G)\mathbf w_i|^2+(\mathbf h_{\text{d},k}^H+ \mathbf h_{\text{r},k}^H  \mathbf \Phi \mathbf G)\mathbf R_0 (\mathbf h_{\text{d},k}+ \mathbf G^{H} \mathbf \Phi^{H} \mathbf h_{\text{r},k}) +\sigma_k^2},
\end{equation}
\begin{equation}\label{eq:SINR_II}
\gamma_k^{\text{II}}=\frac{|(\mathbf h_{\text{d},k}^H+ \mathbf h_{\text{r},k}^H  \mathbf \Phi \mathbf G) \mathbf w_k|^2}{\sum_{i\in\mathcal{K},i\neq k}|(\mathbf h_{\text{d},k}^H+ \mathbf h_{\text{r},k}^H  \mathbf \Phi \mathbf G)\mathbf w_i|^2 +\sigma_k^2}.
\end{equation}
% where $\mathbf h_k = \mathbf h_{\text{d},k}+ \mathbf G^{H} \mathbf \Phi^{H} \mathbf h_{\text{r},k}$ denotes the combined channel vector from the BS to CU $k$. 

\subsection{Sensing Model}
Next, we consider the target sensing. Let $\mathbf H_\text{TRM}$ denote the target response matrix w.r.t. the IRS (i.e., the cascaded channel over the IRS-target-IRS link). The received echo signal by the BS at symbol $t \in \mathcal{T}$ is 
\begin{equation}\label{eq:echo_signal}
\mathbf y(t) = \mathbf G^{T} \mathbf \Phi^{T} \mathbf H_\text{TRM} \mathbf \Phi \mathbf G \mathbf x(t) + \mathbf n_\text{R}(t),
\end{equation}
where $\mathbf n_\text{R}(t)\sim \mathcal{C N}(\mathbf{0}, \sigma_\text{R}^2\mathbf I_M)$ denotes the noise at the BS receiver, which may include the clutter from environment. We stack the transmitted signals, the received signals, and the noise over the radar dwell time as $\mathbf X =[\mathbf x(1),\ldots,\mathbf x(T)]$, $\mathbf Y =[\mathbf y(1),\ldots,\mathbf y(T)]$, and $\mathbf N =[\mathbf n(1),\ldots,\mathbf n(T)]$, respectively. Then, we have
\begin{equation}\label{eq: Y_normal}
\mathbf Y=\mathbf{G}^T\mathbf{\Phi}^{T}\mathbf H_\text{TRM}\mathbf{\Phi}\mathbf{G}\mathbf X+ \mathbf N.
\end{equation}
In particular, we consider two different target models according to the spatial extent of target \cite{xianxin,259642,9652071,huaCRB-RATE,4200705,8579200}.

\subsubsection{Extended Target Case} When the sensing target consists of multiple point-like scatterers in an extended region of space, the echo signals reflected by the target consist of multiple paths from different angles. Normally, the BS has no prior knowledge about the distribution of the scatterers. As a result, the BS first estimates the complete target response matrix $\mathbf H_\text{TRM}$, and then extracts the target parameters from the estimated $\mathbf H_\text{TRM}$ using well-established estimation algorithms such as multiple signal classification (MUSIC) \cite{17564}. As such, we use the CRB for estimating the complete target response matrix $\mathbf H_\text{TRM}$ as the sensing performance metric. We define $\mathbf h = \mathrm{vec}(\mathbf H_\text{TRM}) \in \mathbb{C}^{N^2\times 1}$, and accordingly denote $\bm \zeta=[\mathrm{Re}(\mathbf h^{T}), \mathrm{Im}(\mathbf h^{T})]^{T} \in \mathbb{R}^{2N^2\times 1}$ as the vector of unknown real parameters to be estimated. By vectorizing the stacked echo signal in \eqref{eq: Y_normal}, we have
\begin{equation}\label{eq:vec_data_extended}
\hat{\mathbf y}=\mathrm{vec}(\mathbf Y)=\hat{\mathbf u} + \hat{\mathbf n},
\end{equation}
where $\hat{\mathbf u} = \mathrm{vec}(\mathbf{G}^T\mathbf{\Phi}^{T}\mathbf H_\text{TRM}\mathbf{\Phi}\mathbf{G}\mathbf X)=(\mathbf X^{T}\mathbf{G}^{T}\mathbf{\Phi}^{T}\otimes \mathbf{G}^T\mathbf{\Phi}^{T})\mathbf h$ and $\hat{\mathbf n}=\mathrm{vec}(\mathbf N) \sim \mathcal{C N}(\mathbf{0}, \sigma_\text{R}^2\mathbf I_{MT})$.
Let $\hat{\mathbf F}  \in \mathbb{R}^{2N^2 \times 2N^2}$ denote the Fisher information matrix (FIM) for estimating $\bm \zeta$ from \eqref{eq:vec_data_extended}, each element of which  is given by
\begin{equation}\label{eq:FIM_extended}
\hat{\mathbf F}_{i,j}=\frac{2}{\sigma_\text{R}^2}\mathrm{Re}\left\{\frac{\partial \hat{\mathbf u}^{H}}{\partial \bm \zeta_i}\frac{\partial \hat{\mathbf u}}{\partial \bm \zeta_j}\right\}, i,j\in\{1,\cdots,2N^2\}.
\end{equation}
The FIM for estimating $\bm \zeta$ is \cite{xianxin}
\begin{equation}\label{eq:FIM_partitioned_extended}
\hat{\mathbf F}=
\begin{bmatrix}
\hat{\mathbf F}_{\mathbf h_\text{R} \mathbf h_\text{R}} & \hat{\mathbf F}_{\mathbf h_\text{R} \mathbf h_\text{I}}\\
\hat{\mathbf F}_{\mathbf h_\text{I} \mathbf h_\text{R}} & \hat{\mathbf F}_{\mathbf h_\text{I} \mathbf h_\text{I}}
\end{bmatrix},
\end{equation}
where
\begin{equation}\label{eq:FIM_1_extended}
\hat{\mathbf F}_{\mathbf h_\text{R} \mathbf h_\text{R}}=\hat{\mathbf F}_{\mathbf h_\text{I} \mathbf h_\text{I}}
=\frac{2T}{\sigma_\text{R}^2}\mathrm{Re}\left\{\left(\mathbf{\Phi}^*\mathbf{G}^*\mathbf R_x^{T}\mathbf{G}^{T}\mathbf{\Phi}^{T}\right)\otimes \left(\mathbf{\Phi}^*\mathbf{G}^*\mathbf{G}^{T}\mathbf{\Phi}^{T}\right)\right\},
\end{equation}
\begin{equation}\label{eq:FIM_2_extended}
\hat{\mathbf F}_{\mathbf h_\text{I} \mathbf h_\text{R}}=-\hat{\mathbf F}_{\mathbf h_\text{R} \mathbf h_\text{I}}
=\frac{2T}{\sigma_\text{R}^2}\mathrm{Im}\left\{\left(\mathbf{\Phi}^*\mathbf{G}^*\mathbf R_x^{T}\mathbf{G}^{T}\mathbf{\Phi}^{T}\right)\otimes \left(\mathbf{\Phi}^*\mathbf{G}^*\mathbf{G}^{T}\mathbf{\Phi}^{T}\right)\right\}.
\end{equation}
We consider the CRB for estimating $\bm \zeta$ or  $\mathbf H_\text{TRM}$ as the trace-inverse of the Fisher information matrix $\hat{\mathbf F}$ \cite{xianxin}, i.e., 
\begin{equation}\label{eq:CRB_extended_R_x}
\begin{split}
\text{CRB}_1(\mathbf H_\text{TRM})=&\mathrm{tr}\left(\hat{\mathbf F}^{-1}\right)
 =\frac{\sigma_\text{R}^2}{T}\mathrm{tr}\left(\left(\mathbf{\Phi}\mathbf G\mathbf R_x\mathbf G^{H}\mathbf{\Phi}^{H}\right)^{-1}\right) \mathrm{tr}\left(\left(\mathbf{\Phi}\mathbf G \mathbf G^H\mathbf{\Phi}^{H}\right)^{-1}\right)\\
\stackrel{(a)}{=}&\frac{\sigma_\text{R}^2}{T}\mathrm{tr}\left(\left(\mathbf G\mathbf R_x\mathbf G^{H}\right)^{-1}\right) \mathrm{tr}\left(\left(\mathbf G\mathbf G^{H}\right)^{-1}\right),
\end{split}
\end{equation}
where equality ($a$) holds due to the fact that $\mathbf \Phi \mathbf \Phi^H = \mathbf I_N$. Note that in the extended target case,  $\mathbf H_\text{TRM}$ is estimable only when $\mathrm{rank}(\mathbf G)= N$ and $\mathrm{rank}(\mathbf R_x)\ge N$, since otherwise we have $\text{CRB}_1(\mathbf H_\text{TRM}) \to \infty$ \cite{xianxin}.

\subsubsection{Point Target Case} The point sensing target is modeled as a single scatterer with small spatial extent, and the reflected echo signals only consist of a single path. In this case, the target response matrix $\mathbf H_\text{TRM}$ is modeled as 
\begin{equation}\label{eq:TRM_steering}
\mathbf H_\text{TRM} = \alpha\mathbf a(\theta)\mathbf a^{T} (\theta),
\end{equation}
where $\alpha \in \mathbb{C}$ denotes the complex-valued channel coefficient that depends on the target radar cross section (RCS) and the round-trip path loss of the IRS-target-IRS link, and $\mathbf a(\theta)$ denotes the steering vector at the IRS with angle $\theta$, i.e.,
\begin{equation}\label{eq:steering_vector}
\mathbf a(\theta) = \left[1,e^{j2\pi\frac{d_\text{IRS}\sin \theta}{\lambda}},\cdots,e^{j2\pi\frac{ (N-1)d_\text{IRS}\sin \theta}{\lambda}}\right]^{T},
\end{equation}
with $\theta$ denoting the target DoA w.r.t. the IRS. In \eqref{eq:steering_vector}, $d_\text{IRS}$ denotes the spacing between adjacent reflecting elements at the IRS, and $\lambda$ denotes the carrier wavelength. In this case,  $\bm \xi=[\theta,  \mathrm{Re}\{\alpha\}, \mathrm{Im}\{\alpha\}]^{T} \in \mathbb{R}^{3\times 1}$ denotes the vector of three unknown real parameters to be estimated. Let $\mathbf B= \mathbf b\mathbf b^{T}$ with $\mathbf b= \mathbf{G}^{T}\mathbf{\Phi}^{T}\mathbf a(\theta)$. By vectorizing the stacked echo signal in \eqref{eq: Y_normal}, we have
\begin{equation}\label{eq:vec_data}
\tilde{\mathbf y}=\mathrm{vec}(\mathbf Y)=\tilde{\mathbf u} + \tilde{\mathbf n},
\end{equation}
where $\tilde{\mathbf u} = \alpha \mathrm{vec}(\mathbf B \mathbf X)$ and $\tilde{\mathbf n}=\mathrm{vec}(\mathbf N) \sim \mathcal{C N}(\mathbf{0}, \sigma_\text{R}^2\mathbf I_{MT})$. Let $\tilde{\mathbf F}  \in \mathbb{R}^{3 \times 3}$ denote the FIM for estimating $\bm \xi$ from \eqref{eq:vec_data}. Each element of $\tilde{\mathbf F}$ is given by 
\begin{equation}\label{eq:FIM}
\tilde{\mathbf F}_{i,j}=\frac{2}{\sigma_\text{R}^2}\mathrm{Re}\left\{\frac{\partial \tilde{\mathbf u}^{H}}{\partial \bm \xi_i}\frac{\partial \tilde{\mathbf u}}{\partial \bm \xi_j}\right\}, i,j\in\{1,2,3\}.
\end{equation}
Let $\dot {\mathbf b}$ and $\dot{\mathbf B} = \dot {\mathbf b} \mathbf b^{T} + \mathbf b \dot {\mathbf b}^{T}$ denote the partial derivative of $\mathbf b$ and $\mathbf B$ w.r.t. $\theta$, respectively.
The FIM for estimating $\bm \xi$ is \cite{xianxin}
\begin{equation}\label{eq:FIM_partitioned}
\tilde{\mathbf F}=
\begin{bmatrix}
\tilde{\mathbf{F}}_{\theta \theta} & \tilde{\mathbf{F}}_{\theta \tilde{\bm\alpha}}\\
\tilde{\mathbf{F}}^{T}_{\theta \tilde{\bm\alpha}} & \tilde{\mathbf{F}}_{\tilde{\bm\alpha} \tilde{\bm\alpha}}
\end{bmatrix},
\end{equation}
where
\begin{equation}\label{eq:FIM_1}
\tilde{\mathbf F}_{\theta \theta}=\frac{2T|\alpha|^2}{\sigma_\text{R}^2}\text{tr}\left(\dot {\mathbf B}  \mathbf R_x \dot {\mathbf B}^{H} \right),
\tilde{\mathbf F}_{\theta \tilde{\bm \alpha}}=\frac{2T}{\sigma_\text{R}^2}\mathrm{Re}\left\{\alpha^*\text{tr}\left( \mathbf B\mathbf R_x \dot {\mathbf B}^{H}  \right)[1,j]\right\},
\tilde{\mathbf F}_{\tilde{\bm\alpha} \tilde{\bm\alpha}}=\frac{2T}{\sigma_\text{R}^2}\text{tr}\left( \mathbf B   \mathbf R_x  \mathbf B^{H} \right)\mathbf I_2.
\end{equation}
% \begin{equation}
% \mathbf B(\theta) = \mathbf{G}^{T}\mathbf{\Phi}^{T}\mathbf a(\theta)\mathbf a^T(\theta)\mathbf{\Phi}\mathbf{G},
% \end{equation}
% \begin{equation}\dot{\mathbf B}(\theta) =\frac{\partial \mathbf B(\theta) }{\partial \theta}= j 2\pi \frac{d_\text{IRS}}{\lambda} \cos\theta \mathbf{G}^{T} \mathbf \Phi^T (\mathbf D \mathbf a(\theta)\mathbf a^T(\theta) + \mathbf a(\theta)\mathbf a^T(\theta) \mathbf D)\mathbf \Phi\mathbf{G},
% \end{equation}
% \begin{equation}
% \mathbf D = \mathrm{diag}(0,1,\cdots,N-1).
% \end{equation}
In the point target case, we are interested in estimating the target DoA $\theta$. This is due to the fact that it is difficult to extract the target information from the channel coefficient $\alpha$, as it depends on both the target RCS and the distance-dependent path loss of the IRS-target-IRS link, which are usually unknown \cite{9652071,xianxin,huaCRB-RATE}. The CRB for estimating the target DoA $\theta$ equals to the first diagonal element of $\tilde{\mathbf F}^{-1}$ \cite{xianxin}, i.e., 
% When the BS have a-prior information about the target potential directions (e.g., in the target tracking
% scenario), we are particularly interested in the sensing performance within a certain region or a set of directions, denoted by $\theta_1, \cdots, \theta_L$. Let $\mathcal L \triangleq \{1,\cdots, L\}$ denote the set of desired sensing angles. In this case, we use the worst-case CRB for estimating the target DoA in the potential angles as the sensing performance measure. Based on the derivation in \cite{xianxin}, the CRB for estimating the target DoA at angle $\theta$ is given in the following:
\begin{subequations}\label{eq:CRB_point}
\begin{align}\label{eq:CRB_point_1}
\text{CRB}_2(\theta)=& [\tilde{\mathbf F}^{-1}]_{1,1}\! =\![\tilde{\mathbf F}_{\theta \theta}-\tilde{\mathbf F}_{\theta \tilde{\bm\alpha}}\tilde{\mathbf F}_{\tilde{\bm\alpha} \tilde{\bm\alpha}}^{-1}\tilde{\mathbf F}_{\theta \tilde{\bm\alpha}}^{T}]^{-1}
=\frac{\sigma_\text{R}^2}{2T|\alpha|^2\left(\mathrm{tr}\left(\dot {\mathbf B} \mathbf R_x \dot {\mathbf B}^{H}\right)-\frac{|\mathrm{tr}\left(\mathbf B \mathbf R_x  \dot {\mathbf B}^{H}\right)|^2}{\mathrm{tr}\left(\mathbf B \mathbf R_x \mathbf B^{H}\right)}\right)}\\\label{eq:CRB_point_2}
% =&\frac{\sigma_\text{R}^2}{2T|\alpha|^2\left(\mathbf b^{H}\mathbf R_x^* \mathbf b\left(\|\dot{\mathbf c}\|^2-\frac{|\dot{\mathbf c}^{H}\mathbf c|^2}{\|\mathbf c\|^2}\right)+\|\mathbf c\|^2\left(\dot{\mathbf b}^{H}\mathbf R_x^* \dot{\mathbf b}-\frac{|\dot{\mathbf b}^{H}\mathbf R_x^*\mathbf b|^2}{\mathbf b^{H}\mathbf R_x^* \mathbf b}\right)\right)}\\
=&\frac{\sigma_\text{R}^2 \lambda^2/\left(8T|\alpha|^2\pi^2d_\text{IRS}^2\cos^2(\theta)\right)}{\mathbf v^{H} \mathbf{R}_{2}  \mathbf v \left(\mathbf v^{H} \mathbf D \mathbf{R}_{1}  \mathbf D\mathbf v-\frac{|\mathbf v^{H} \mathbf D \mathbf{R}_{1} \mathbf v|^2}{\mathbf v^{H} \mathbf{R}_{1} \mathbf v}\right) + \mathbf v^{H} \mathbf{R}_{1} \mathbf v \left(\mathbf v^{H} \mathbf D \mathbf{R}_{2} \mathbf D \mathbf v-\frac{|\mathbf v^{H} \mathbf D \mathbf{R}_{2} \mathbf v|^2}{\mathbf v^{H} \mathbf{R}_{2} \mathbf v}\right)},
\end{align}
\end{subequations}
where $\mathbf{R}_{1}= \mathrm{diag}\left(\mathbf a^{H}(\theta)\right)\mathbf{G}^* \mathbf{G}^{T}\mathrm{diag}\left(\mathbf a(\theta)\right)$, $\mathbf{R}_{2}= \mathrm{diag}\left(\mathbf a^{H}(\theta)\right)\mathbf{G}^* \mathbf R^*_x \mathbf{G}^{T}\mathrm{diag}\left(\mathbf a(\theta)\right)$, and $\mathbf D = \mathrm{diag}(0,1, \cdots ,N-1)$. Note that the CRB formulas in \eqref{eq:CRB_point_1} and \eqref{eq:CRB_point_2}  are expressed in different forms to facilitate the transmit beamforming and reflective beamforming optimization, respectively. 
% \begin{subequations}
% \begin{align}
% % \mathbf B  =&  \mathbf G^{T}\mathrm{diag}(\mathbf a(\theta))\mathbf{v} \mathbf v^{T}\mathrm{diag}(\mathbf a^{T}(\theta))\mathbf G,\\
% % \dot {\mathbf B} =& j 2\pi \frac{d_\text{IRS}}{\lambda} \cos\theta \mathbf{G}_t^{T}\mathrm{diag}(\mathbf a(\theta))(\mathbf D\mathbf{v} \mathbf v^{T}+ \mathbf{v} \mathbf v^{T}\mathbf D^{T})\mathrm{diag}(\mathbf a^{T}(\theta))\mathbf G_t,\\
% % \mathbf D =& \mathrm{diag}(0,1,\cdots,N-1),\\
% \mathbf{R}_{1}=& \mathrm{diag}(\mathbf a^{H})\mathbf{G}^* \mathbf{G}^{T}\mathrm{diag}(\mathbf a(\theta)),\\
% \mathbf{R}_{2}=& \mathrm{diag}(\mathbf a^{H})\mathbf{G}^* \mathbf R^*_x \mathbf{G}^{T}\mathrm{diag}(\mathbf a(\theta)).
% \end{align}
% \end{subequations}
% For notational convenience, in the sequel we drop $\theta$ in $\mathbf B(\theta)$, $\dot{\mathbf B}(\theta)$, $\mathbf{R}_{1}(\theta)$, and $\mathbf{R}_{2}(\theta)$, and accordingly denote them as $\mathbf B$, $\dot{\mathbf B}$, $\mathbf{R}_{1}$, and $\mathbf{R}_{2}$, respectively.

\subsection{Problem Formulation}
We aim to minimize the CRB for target estimation (i.e., $\text{CRB}_1(\mathbf H_\text{TRM})$ for the extended target case and $\text{CRB}_2(\theta)$ for the point target case), by jointly optimizing the transmit beamformers $\{\mathbf w_k\}$ and $ \mathbf R_0$ at the BS and the reflective beamformer $\mathbf v$ or $\mathbf \Phi$ at the IRS, subject to the minimum SINR constraints at individual CUs and the maximum transmit power constraint at the BS.
\subsubsection{Problem Formulation with Extended Target}
For the extended target case, we use $\text{CRB}_1(\mathbf H_\text{TRM})$ in \eqref{eq:CRB_extended_R_x} as the sensing performance metric. Based on the CRB given in \eqref{eq:CRB_extended_R_x}, minimizing $\text{CRB}_1(\mathbf H_\text{TRM})$ is equivalent to minimizing 
$\mathrm{tr}\left(\left(\mathbf G\left(\sum_{k\in \mathcal{K}}\mathbf w_k \mathbf w_k^H + \mathbf R_0\right)\mathbf G^H\right)^{-1}\right)$. As a result, the SINR-constrained CRB minimization problems with Type-I and Type-II CU receivers are formulated as (P1) and (P2), respectively. 
\begin{subequations}
\begin{align} \nonumber
  &\qquad \quad\text{(P1)}:     \min_{\{\mathbf w_k\}, \mathbf R_0, \mathbf \Phi}  \quad    \mathrm{tr}\left(\left(\mathbf G\left(\sum_{k\in \mathcal{K}}\mathbf w_k \mathbf w_k^H + \mathbf R_0\right)\mathbf G^H\right)^{-1}\right) \\  \label{eq:SINR_minimum_I}
   &\text{s.t.}~ \frac{|(\mathbf h_{\text{d},k}^H+ \mathbf h_{\text{r},k}^H  \mathbf \Phi \mathbf G) \mathbf w_k|^2}{\sum_{i\in\mathcal{K},i\neq k}\!|(\mathbf h_{\text{d},k}^H \!+\! \mathbf h_{\text{r},k}^H  \mathbf \Phi \mathbf G)\mathbf w_i|^2+(\mathbf h_{\text{d},k}^H\!+\!  \mathbf h_{\text{r},k}^H  \mathbf \Phi \mathbf G)\mathbf R_0 (\mathbf h_{\text{d},k}\!+\! \mathbf G^{H} \mathbf \Phi^{H} \mathbf h_{\text{r},k}) +\sigma_k^2}\ge \Gamma_k, \forall k\in \!\mathcal{K}\\ \label{eq:power}
  &\quad ~ \sum_{k\in\mathcal{K}}\|\mathbf w_k\|^2  + \mathrm {tr}(\mathbf R_0) \le P_0 \\\label{eq:semi}
   &\quad ~\mathbf R_0 \succeq \mathbf{0}\\\label{eq:phase_1}
  &\quad ~ \mathbf |\mathbf \Phi_{n,n}|=1, \forall n\in \mathcal{N}.
  \end{align}
\end{subequations}
\begin{subequations}
\begin{align} \nonumber
  \text{(P2)}:    \nonumber \min_{\{\mathbf w_k\}, \mathbf R_0, \mathbf \Phi}&\quad    \mathrm{tr}\left(\left(\mathbf G\left(\sum_{k\in \mathcal{K}}\mathbf w_k \mathbf w_k^H + \mathbf R_0\right)\mathbf G^H\right)^{-1}\right)  \\   \label{eq:SINR_minimum_II}
   \text{s.t.}&  \quad  \frac{|(\mathbf h_{\text{d},k}^H+ \mathbf h_{\text{r},k}^H  \mathbf \Phi \mathbf G) \mathbf w_k|^2}{\sum_{i\in\mathcal{K},i\neq k}|(\mathbf h_{\text{d},k}^H+ \mathbf h_{\text{r},k}^H  \mathbf \Phi \mathbf G)\mathbf w_i|^2 +\sigma_k^2}\ge \Gamma_k, \forall k\in \mathcal{K}\tag{24}\\ 
  & \nonumber  \quad \eqref{eq:power}, ~\eqref{eq:semi},~\text{and}~\eqref{eq:phase_1}.
\end{align}
\end{subequations}
Notice that in problems (P1) and (P2), the SINR constraints in \eqref{eq:SINR_minimum_I} and \eqref{eq:SINR_minimum_II} as well as the unit-modulus constraints on the reflecting coefficients in \eqref{eq:phase_1} are non-convex, and the transmit and reflective beamformers are coupled. Therefore, problems (P1) and (P2) are non-convex and thus challenging to solve. We will address these two problems in Section III. Furthermore, by comparing problems (P1) versus (P2), it is clear that the optimal CRB value of (P2) serves as a lower bound on that of (P1), which is due to the fact that every feasible solution to (P1) is also feasible for (P2) but not vice versa. This shows the benefit of sensing signal interference cancellation at Type-II CU receivers.

\subsubsection{Problem Formulation with Point Target}
For the point target case, we use $\text{CRB}_2(\theta)$ in \eqref{eq:CRB_point} as the sensing performance metric. We assume that the BS roughly knows the information of $\theta$ to implement the joint beamforming design, which corresponds to the target tracking scenario in practice. As a result, the SINR-constrained CRB minimization problems with Type-I and Type-II CU receivers are formulated as (P3) and (P4), respectively. 
\begin{subequations}
\begin{align} \nonumber
  \text{(P3)}:  \min_{\{\mathbf w_k\}, \mathbf R_0, \mathbf \Phi} & \quad  \frac{\sigma_\text{R}^2}{2T|\alpha|^2\left(\mathrm{tr}\left(\dot {\mathbf B} \left(\sum_{k\in\mathcal{K}} \mathbf w_k\mathbf w_k^{H} + \mathbf R_0\right) \dot {\mathbf B}^{H}\right)-\frac{\left|\mathrm{tr}\left(\mathbf B \left(\sum_{k\in\mathcal{K}} \mathbf w_k\mathbf w_k^{H} + \mathbf R_0\right) \dot {\mathbf B}^{H}\right)\right|^2}{\mathrm{tr}\left(\mathbf B \left(\sum_{k\in\mathcal{K}} \mathbf w_k\mathbf w_k^{H} + \mathbf R_0\right) \mathbf B^{H}\right)}\right)} \\  
   \nonumber\text{s.t.}&  \quad \eqref{eq:SINR_minimum_I},~\eqref{eq:power}, ~\eqref{eq:semi},~\text{and}~\eqref{eq:phase_1}.
\end{align}
\end{subequations}
\begin{subequations}
\begin{align} \nonumber
  \text{(P4)}:  \min_{\{\mathbf w_k\}, \mathbf R_0, \mathbf \Phi}& \quad  \frac{\sigma_\text{R}^2}{2T|\alpha|^2\left(\mathrm{tr}\left(\dot {\mathbf B} \left(\sum_{k\in\mathcal{K}} \mathbf w_k\mathbf w_k^{H} + \mathbf R_0\right) \dot {\mathbf B}^{H}\right)-\frac{\left|\mathrm{tr}\left(\mathbf B \left(\sum_{k\in\mathcal{K}} \mathbf w_k\mathbf w_k^{H} + \mathbf R_0\right) \dot {\mathbf B}^{H}\right)\right|^2}{\mathrm{tr}\left(\mathbf B \left(\sum_{k\in\mathcal{K}} \mathbf w_k\mathbf w_k^{H} + \mathbf R_0\right) \mathbf B^{H}\right)}\right)} \\  
   \nonumber \text{s.t.}& \quad \eqref{eq:SINR_minimum_II},~\eqref{eq:power}, ~\eqref{eq:semi},~\text{and}~\eqref{eq:phase_1}.
\end{align}
\end{subequations}
Notice that in problems (P3) and (P4), besides the non-convex constraints in \eqref{eq:SINR_minimum_I}, \eqref{eq:SINR_minimum_II}, and \eqref{eq:phase_1}, the objective functions are both non-convex. Therefore, problems (P3) and (P4) are non-convex and more difficult to solve than (P1) and (P2). We will address these two problems in Section IV. Similarly as for (P1) and (P2), thanks to the  sensing signal interference cancellation at Type-II CU receivers, the optimal CRB value of problem (P4) is a lower bound of that of (P3). 

\section{Joint Beamforming Solutions to (P1) and (P2) with Extended Target}
In this section, we develop efficient algorithms based on the alternating optimization technique to solve the non-convex SINR-constrained CRB minimization problems (P1) and (P2) with extended target, in which the transmit beamformers $\{\mathbf w_k\}$ and $\mathbf R_0$ at the BS and the reflective beamformer $\mathbf \Phi$ at the IRS are iteratively optimized by using optimization techniques such as SDR. As problems (P1) and (P2) have similar structures, we first focus on solving (P1) in Section~III-A, and then discuss the solution to (P2)  in Section III-B by addressing its difference from (P1). To gain more insights, Section III-C further analyzes the solution structures of problems (P1) and (P2).
%considers a special case with one single CU  analyzes the performance by comparing with the case without dedicated sensing signals.

% Problem (P1) is difficult to solve due to the non-convexity of the objective function, the unit-modulus constraint in \eqref{eq:phase_1}, and the coupled relationship between the transmit and reflective beamformers. In order to solve the non-convex problem (P1) efficiently, we use an alternating optimization technique by decomposing (P1) into two subproblems and iteratively solving them in an alternating manner. For each subproblem, we use various optimization techniques such as SCA and SDR. 

\subsection{Proposed Solution to (P1) with Type-I CU Receivers} 

\subsubsection{Transmit Beamforming Optimization for (P1)}
First, we optimize the transmit beamformers $\{\mathbf w_k\}$ and $\mathbf R_0$ in (P1) under any given reflective beamformer $\mathbf \Phi$. 
 The transmit beamforming optimization problem is formulated as
\begin{subequations}
\begin{align} \nonumber
  \text{(P1.1)}:
       \min_{\{\mathbf w_k\}, \mathbf R_0}& \quad   \mathrm{tr}\left(\left(\mathbf G\left(\sum_{k\in\mathcal{K}} \mathbf w_k\mathbf w_k^{H} + \mathbf R_0\right)\mathbf G^{H}\right)^{-1}\right)   \\ \nonumber
  \text{s.t.} & \quad   \eqref{eq:SINR_minimum_I}, ~ \eqref{eq:power},~\text{and}~\eqref{eq:semi}.
\end{align}
\end{subequations}

We use the SDR technique to obtain the {\it optimal} solution to problem (P1.1). Define $\mathbf  W_k = \mathbf  w_k \mathbf  w_k^{H}$, with $\mathbf  W_k \succeq \mathbf{0}$ and $\mathrm{rank}(\mathbf  W_k) \le 1, \forall k\in \mathcal K$. Let $\mathbf h_k = \mathbf h_{\text{d},k}+ \mathbf G^{H} \mathbf \Phi^{H} \mathbf h_{\text{r},k}$ denote the combined channel vector from the BS to CU $k$. By defining  $\mathbf  H_k = \mathbf  h_{k}\mathbf  h_{k}^{H}, \forall k \in \mathcal{K}$, problem (P1.1) is equivalently reformulated as 
\setcounter{equation}{24}
\begin{subequations}
\begin{align} \nonumber 
  \text{(P1.2)}:  \min_{\{\mathbf W_k\}, \mathbf R_0}& \quad   \mathrm{tr}\left(\left(\mathbf G\left(\sum_{k\in\mathcal{K}} \mathbf W_k + \mathbf R_0\right)\mathbf G^{H}\right)^{-1}\right)  \\  
  \text{s.t.}&  \quad  
  %\label{eq: SINR_W_I} \frac{1}{\Gamma_k}\mathrm{tr}(\mathbf  H_k\mathbf  W_k)-\sum_{i\in\mathcal{K},i\neq k}\mathrm{tr}(\mathbf  H_k\mathbf  W_i)-\mathrm{tr}(\mathbf  H_k\mathbf  R_0) \ge \sigma^2_k, \forall k \in \mathcal{K}\\
  \label{eq: SINR_W_I_reformulated} (1+\frac{1}{\Gamma_k})\mathrm{tr}(\mathbf  H_k\mathbf  W_k)-\mathrm{tr}\left(\mathbf  H_k \left(\sum_{k\in\mathcal{K}}\mathbf  W_k+\mathbf  R_0\right)\right) \ge \sigma^2_k, \forall k \in \mathcal{K}\\
  \label{eq: power_W}&  \quad \sum_{k\in\mathcal{K}}\mathrm {tr}(\mathbf W_k) + \mathrm {tr}(\mathbf R_0)  \le P_0 \\ 
  \label{eq:W_semi}&  \quad \mathbf R_0 \succeq \mathbf{0}, \mathbf  W_k \succeq \mathbf{0}, \forall k \in \mathcal{K}\\
  \label{eq:rank-one_W}& \quad \mathrm{rank}(\mathbf W_k) \le 1, \forall k \in \mathcal{K}.
\end{align}
\end{subequations}

Next, we drop the rank-one constraints in \eqref{eq:rank-one_W} to get the SDR version of (P1.2), denoted by (SDR1.2), which is a convex semi-definite program (SDP) and thus can be optimally solved by convex solvers such as CVX \cite{cvx}. 
%Next, we reformulate problem (SDR1.2) to reduce the computation complexity based on the following proposition. The SINR constraint in \eqref{eq: SINR_W_I} is equivalently rewritten as
% \begin{equation}\label{eq: SINR_W_I_reformulated}
%  (1+\frac{1}{\Gamma_k})\mathrm{tr}(\mathbf  H_k\mathbf  W_k)-\mathrm{tr}\left(\mathbf  H_k \left(\sum_{k\in\mathcal{K}}\mathbf  W_k+\mathbf  R_0\right)\right) \ge \sigma^2_k, \forall k \in \mathcal{K}.
% \end{equation}
Let $\{\mathbf W^\star_k\}$ and  $\mathbf R^\star_0$ denote the optimal solution to (SDR1.2). We then have the following proposition. 
\begin{proposition} \label{prop:SDR}
The SDR of (P1.2) or equivalently (P1.1) is tight, i.e., problems (P1.1), (P1.2), and (SDR1.2) have the same optimal value.  Given the optimal solution $\{\mathbf W_k^{\star}\}$ and $\mathbf R_0^{\star}$ to (SDR1.2), the optimal solution to (P1.1) is
\begin{equation} \label{eq:w_new}
\mathbf w_k^\text{opt,I} = (\mathbf h_k^{H} \mathbf W_k^{\star} \mathbf h_k)^{-1/2}\mathbf W_k^{\star} \mathbf h_k, \forall k \in\mathcal{K},
\end{equation}
 \begin{equation} \label{eq:R_new}
\mathbf R_0^\text{opt,I} = \mathbf R_0^{\star} + \sum_{k\in \mathcal{K}} \mathbf W_k^{\star}- \sum_{k\in \mathcal{K}}\mathbf w_k^\text{opt,I}(\mathbf w_k^\text{opt,I})^H.
\end{equation}
\end{proposition}
\begin{IEEEproof}
See Appendix~\ref{sec:proof_of_proposition_SDR}.
\end{IEEEproof}
Therefore, the optimal solution to the transmit beamforming problem (P1.1) is obtained. To gain more insights, we further present the following proposition, which shows the solution structure of (SDR1.2).
\begin{proposition}\label{proposition_zero}
There exists one optimal solution to problem (SDR2.1), such that $\mathbf R_0^\star = \mathbf 0$.
\end{proposition}
\begin{IEEEproof}
See Appendix~\ref{sec:proof_of_proposition_zero}.
\end{IEEEproof}
However, the property in Proposition~\ref{proposition_zero} may not hold for problem (P1.2) or (P1.1), i.e., there may not exist one optimal solution for them with $\mathbf R_0^\text{opt,I}=\mathbf 0$ based on \eqref{eq:R_new}. In particular, when the number of CUs $K$ is less than $N$, we have $\mathrm{rank}\left(\sum_{k\in \mathcal K} \mathbf w_k \mathbf w_k^H\right) < N$, and as a result, we need the dedicated sensing signal with non-zero $\mathbf R_0^\text{opt,I}$ to ensure $\mathrm{rank}(\mathbf R_x) \ge N$ for making the extended target estimation feasible or making $\text{CRB}_1(\mathbf H_\text{TRM})$ finite. This thus shows the solution structure differences between problems (P1.1) or (P1.2) versus (SDR1.2).

\subsubsection{Reflective Beamforming Optimization for (P1)}
Next, we optimize the reflective beamformer $\mathbf \Phi$ in (P1) under any given transmit beamformers $\{\mathbf w_k\}$ and $\mathbf R_0$. As $\text{CRB}_1(\mathbf H_\text{TRM})$ in \eqref{eq:CRB_extended_R_x} is independent of the reflective beamformer $\mathbf \Phi$, (P1) is simplified as the following feasibility problem. 
\begin{subequations}
\begin{align} \nonumber
  \text{(P1.3)}:     \text{Find}&  \  \ \mathbf \Phi  \\ \nonumber
  \text{s.t.} & \ \  \eqref{eq:SINR_minimum_I}~\text{and}~\eqref{eq:phase_1}.
\end{align}
\end{subequations}
Motivated by the design in \cite{8811733}, we further transform (P1.3) into the following optimization problem (P1.4) with an explicit objective for increasing the SINR at all CUs. This is expected to help achieve a better converged solution.
\setcounter{equation}{27}
\begin{subequations}
\begin{align} \nonumber
  \text{(P1.4)}:     \max_{\mathbf \Phi,\{\beta_k\}}&  \  \ \sum_{k\in\mathcal{K}}\beta_k  \\  \nonumber
  \text{s.t.}&  \quad |(\mathbf h_{\text{d},k}^H+ \mathbf h_{\text{r},k}^H  \mathbf \Phi \mathbf G) \mathbf w_k|^2-\Gamma_k\sum_{i\in\mathcal{K},i\neq k}|(\mathbf h_{\text{d},k}^H+ \mathbf h_{\text{r},k}^H  \mathbf \Phi \mathbf G)\mathbf w_i|^2 \\
  &\quad - \Gamma_k(\mathbf h_{\text{d},k}^H+ \mathbf h_{\text{r},k}^H  \mathbf \Phi \mathbf G)\mathbf R_0 (\mathbf h_{\text{d},k}+ \mathbf G^{H} \mathbf \Phi^{H} \mathbf h_{\text{r},k}) -\Gamma_k\sigma_k^2 \ge \beta_k, \forall k \in \mathcal{K} \\
  &\quad \beta_k \ge 0, \forall k \in \mathcal{K}\\ 
  & \quad \mathbf |\mathbf \Phi_{n,n}|=1, \forall n\in \mathcal{N}.
\end{align}
\end{subequations}
Note that (P1.4) has a similar structure as the reflective beamforming design problem in \cite[(48)]{8811733} for IRS-assisted wireless communications, which can be solved by using the SDR technique together with Gaussian randomization \cite{8811733}, for which the details are omitted for brevity.

\begin{table}
\centering{
\caption{Proposed Algorithm for Solving Problem (P1)\label{tab:table1}}}
\vspace{-0.5cm}
 \hrule
\vspace{0.1cm} \textbf{Algorithm 1}  \vspace{0.1cm}
\hrule 
        \begin{enumerate}[a)]
            \item Set iteration index $l= 1$ and initialize  $\mathbf{\Phi}^{(l)}$ randomly.
            \item \textbf{Repeat}: \begin{enumerate}[1)]
            				\item Under given $\mathbf{\Phi}^{(l)}$, solve problem (SDR1.2) to obtain the optimal solution as $\{{\mathbf W}_k^{\star(l)}\}$ and ${\mathbf R}_0^{\star(l)}$.
            				\item Construct the optimal rank-one solution $\{\mathbf w_k^{(l)}\}$ and $ \mathbf R_0^{(l)}$ to (P1.1) based on $\{{\mathbf W}_k^{\star(l)}\}$ and $ {\mathbf R}_0^{\star(l)}$ by using Proposition~\ref{prop:SDR}.
            				\item Under given $\{\mathbf w_k^{(l)}\}$ and $ \mathbf R_0^{(l)}$, solve problem (P1.4) to obtain the reflective beamformer $\mathbf{\Phi}^{(l+1)}$.
            				\item Update $l\gets l+1$.
            				\end{enumerate}
            				\item \textbf{Until} the convergence criterion is met or the maximum number of outer iterations is reached.
        \end{enumerate}
\vspace{0.1cm} \hrule\vspace{-20pt}\label{algorithm:new0}
\end{table}

\subsubsection{Complete Algorithm for Solving (P1)}
By combining the transmit and reflective beamforming designs in Sections III-A-(1) and III-A-(2), together with the alternating optimization, we have the complete algorithm to solve (P1), which is summarized as Algorithm~1 in Table~I. Notice that in each iteration of Algorithm~1, (P1.1) is optimally solved, which leads to a non-increasing CRB value, and (P1.4) does not change the CRB value but leads to a larger feasible set for the transmit beamforming design. As a result, the convergence of Algorithm~1 for solving problem (P1) is ensured.

\subsection{Proposed Solution to Problem (P2) with Type-II CU Receivers}
In this subsection, we consider problem (P2) with Type-II CU receivers, which can be solved similarly as Algorithm~1 for (P1) based on  alternating optimization. Therefore, we present the transmit and reflective beamforming design in the following briefly.

First, we optimize the transmit beamformers $\{\mathbf w_k\}$ and $\mathbf R_0$ in (P2) with any given reflective beamformer $\mathbf \Phi$, which is given by 
\begin{subequations}
\begin{align} \nonumber
  \text{(P2.1)}:
       \min_{\{\mathbf w_k\}, \mathbf R_0}& \quad   \mathrm{tr}\left(\left(\mathbf G\left(\sum_{k\in\mathcal{K}} \mathbf w_k\mathbf w_k^{H} + \mathbf R_0\right)\mathbf G^{H}\right)^{-1}\right)   \\ \nonumber
  \text{s.t.} & \quad   \eqref{eq:SINR_minimum_II},~ \eqref{eq:power}, ~\text{and}~\eqref{eq:semi}.
\end{align}
\end{subequations}
Problem (P2.1) can be optimally solved by using SDR similarly as for (P1.1). Towards this end, we express the SDR of (P2.1) (by introducing $\mathbf W_k = \mathbf w_k\mathbf w_k^H, \forall k\in \mathcal K$ and dropping the rank-one constraints on $\{\mathbf W_k\}$) as (SDR2.1), which is a convex SDP that can be optimally solved via CVX. 
\begin{subequations}
\begin{align} \nonumber 
  \text{(SDR2.1)}:  \min_{\{\mathbf W_k\}, \mathbf R_0}& \quad    \mathrm{tr}\left(\left(\mathbf G\left(\sum_{k\in\mathcal{K}} \mathbf W_k+ \mathbf R_0\right)\mathbf G^{H}\right)^{-1}\right)  \\  
  \text{s.t.}&  \quad \label{eq: SINR_W_II} \frac{1}{\Gamma_k}\mathrm{tr}(\mathbf  H_k\mathbf  W_k)\!-\hspace{-0.5em} \sum_{i\in\mathcal{K},i\neq k}\mathrm{tr}(\mathbf  H_k\mathbf  W_i)\ge \sigma^2_k, \forall k \in \mathcal{K} \tag{29}\\\nonumber
  & \quad \eqref{eq: power_W}~\text{and}~\eqref{eq:W_semi}.
\end{align}
\end{subequations}
Let $\{\mathbf W_k^{\star\star}\}$ and $\mathbf R_0^{\star\star}$ denote the optimal solution to (SDR2.1). We have the following proposition.
\begin{proposition} \label{prop:SDR_2}
The SDR of (P2.1) is tight, i.e., problems (P2.1) and (SDR2.1) have the same optimal value.  Given the optimal solution $\{\mathbf W_k^{\star\star}\}$ and $\mathbf R_0^{\star\star}$ to (SDR2.1), the optimal solution to (P2.1) is
% Problems (P2.1) and (SDR2.1) have the same optimal value. The  optimal solution $\{\mathbf w_k^\text{opt,II}\}$ and $\mathbf R_0^\text{opt,II}$ to problem (P2.1) can be constructed based on $\{\bar{\mathbf W}_k\}$ and $\bar{\mathbf R}_0$ to (SDR2.1) similarly as in \eqref{eq:w_new} and \eqref{eq:R_new}, i.e.,
\setcounter{equation}{29}
\begin{equation} \label{eq:w_new_2}
\mathbf w_k^\text{opt,II} = (\mathbf h_k^{H} \mathbf W_k^{\star\star} \mathbf h_k)^{-1/2}\mathbf W_k^{\star\star} \mathbf h_k, \forall k \in \mathcal{K},
\end{equation}
 \begin{equation} \label{eq:R_new_2}
\mathbf R_0^\text{opt,II} = \mathbf R_0^{\star\star}+  \sum_{k\in \mathcal{K}} \mathbf W_k^{\star\star} - \sum_{k\in \mathcal{K}}\mathbf w_k^\text{opt,II}(\mathbf w_k^\text{opt,II})^H.
\end{equation}
\end{proposition}
\begin{IEEEproof}
See Appendix \ref{sec:proof_of_proposition_1_2}.
\end{IEEEproof}
It is worth noting that at the optimality, it generally holds that $\mathbf R_0^{\star\star} \neq \mathbf 0$ for problem (SDR2.1) and $\mathbf R_0^\text{opt,II} \neq \mathbf 0$ for problem (P2.1). This is different from problem (SDR1.2) and means that the dedicated sensing signals are generally needed when Type-II CU receivers are considered.

Next, we optimize the reflective beamformer $\mathbf \Phi$ in (P2) with any given transmit beamformers $\{\mathbf w_k\}$ and $\mathbf R_0$, for which the optimization problem becomes
\begin{subequations}
\begin{align} \nonumber
  \text{(P2.2)}:     \text{Find}&  \  \ \mathbf \Phi  \\ \nonumber
  \text{s.t.} & \ \  \eqref{eq:SINR_minimum_II}~\text{and}~\eqref{eq:phase_1}.
\end{align}
\end{subequations}
As problem (P2.2) has a similar structure as (P1.3), it can be solved similarly by using the algorithm in Section III-A-(2), for which the details are omitted. Therefore, by combining the solutions to (P2.1) and (P2.2) together with alternating optimization, problem (P2) is finally solved. 

\subsection{Solution Structure}
In this subsection, we analyze the solution structures of problems (P1) and (P2) to gain more insights. First, we consider the special case with one single CU. In this case, we have the following two propositions. 
\begin{proposition} \label{prop:single_user_Type-I}
At the optimality of (P1) with Type-I CU receivers, when $K=1$, the dedicated sensing beamformers should lie in the null space of the channel vector $\mathbf h_1$ from the BS to the CU, i.e., $\mathbf h_1^H \mathbf R_0^\text{opt,I} \mathbf h_1=0$.
\end{proposition} 
\begin{IEEEproof}
This proposition follows directly from Proposition~\ref{prop:SDR} and Proposition~\ref{proposition_zero}.
\end{IEEEproof}
% Present Proposition 3 here 
% \textcolor{blue}{It is also interesting to discuss the solution structure of (P1.2)/(P1.1) in the special case with $K =1$ CU.
% Based on Proposition~\ref{prop:SDR} and Proposition~\ref{proposition_zero}, in this case when $K=1$, we have $\mathbf h_1^H \mathbf R_0^\text{opt,I} \mathbf h_1=0$. Therefore, the optimal dedicated sensing beamformers should lie in  the null space of the channel vector $\mathbf h_1$ from the BS to the CU.}
\begin{proposition} \label{prop:single-user}
When $K=1$, problems (P1) and (P2) become identical.
\end{proposition} 
\begin{IEEEproof}
Based on Proposition~\ref{prop:single_user_Type-I}, at the optimality with Type-I CU receiver, the dedicated sensing signals with covariance matrix $\mathbf R_0$ will not introduce interference towards the CU. As such, the interference cancellation by Type-II CU receiver cannot provide any SINR gain over the Type-I CU receiver. Thus, (P1.1) and (P2.1) have the same optimal objective value under any identical reflective beamformer $\mathbf \Phi$. As a result, problems (P1) and (P2) become identical.
\end{IEEEproof}

Propositions \ref{prop:single_user_Type-I} and \ref{prop:single-user} show that for the special case with one single CU, Type-II CU receivers do not provide any performance gain over their Type-I counterparts. 

Next, we discuss the solution structure of $\mathbf R_0$ for problems (P1) and (P2). Notice that based on \eqref{eq:CRB_extended_R_x}, in order for the $\text{CRB}_1(\mathbf H_\text{TRM})$ to be bounded from above, it must follow that 
\setcounter{equation}{31}
\begin{equation}
\mathrm{rank}\left(\mathbf R_x\right)=\mathrm{rank}\left(\sum_{k\in\mathcal{K}} \mathbf w_k\mathbf w_k^{H} + \mathbf R_0\right)\ge N.
\end{equation}
As a result, if $K<N$, then $\mathbf R_0$ should be non-zero as the dedicated sensing signal is needed to ensure the CRB to be bounded from above. By contrast, when $K \ge N$, it is observed from extensive simulations that $\mathbf R_0^\text{opt,I}= \mathbf 0$ occurs for problem (P1) but $\mathbf R_0^\text{opt,II} \neq \mathbf 0$ occurs for problem (P2). This shows that dedicated sensing signals are useful only for Type-II CU receivers when the sensing signal interference cancellation is implemented.

\section{Joint Beamforming Solutions to (P3) and (P4) with Point Target}
In this section, we develop efficient algorithms based on the alternating optimization technique to solve the non-convex SINR-constrained CRB minimization problems (P3) and (P4) with point target, in which the transmit beamformers $\{\mathbf w_k\}$ and $\mathbf R_0$ at the BS and the reflective beamformer $\mathbf v$ at the IRS are iteratively optimized by using optimization techniques including SDR and SCA. As problems (P3) and (P4) have similar structures, we first consider solving (P3) in Section IV-A, and then discuss the solution to (P4) in Section IV-B by highlighting its difference from (P3). To gain more insights, Section IV-C further analyzes the solution structures of problems (P3) and (P4).

\subsection{Proposed Solution to (P3) with Type-I Receivers} 

\subsubsection{Transmit Beamforming Optimization for (P3)}
First, we optimize the transmit beamformers $\mathbf R_0$ and $\{\mathbf w_k\}$ in (P3) under any given reflective beamformer $\mathbf v$, in which the CRB expression in \eqref{eq:CRB_point_1} is used. In this case, the transmit beamforming optimization problem is formulated as
\begin{subequations}
\begin{align} \nonumber
  \text{(P3.1)}: 
  \max_{\{\mathbf w_k\}, \mathbf R_0} & \quad  \mathrm{tr}\left(\dot {\mathbf B} \left(\sum_{k\in\mathcal{K}} \mathbf w_k\mathbf w_k^{H} + \mathbf R_0\right) \dot {\mathbf B}^{H}\right)-\frac{\left|\mathrm{tr}\left(\mathbf B \left(\sum_{k\in\mathcal{K}} \mathbf w_k\mathbf w_k^{H} + \mathbf R_0\right) \dot {\mathbf B}^{H}\right)\right|^2}{\mathrm{tr}\left(\mathbf B \left(\sum_{k\in\mathcal{K}} \mathbf w_k\mathbf w_k^{H} + \mathbf R_0\right) \mathbf B^{H}\right)}\\ \notag 
   \text{s.t.}& \quad \eqref{eq:SINR_minimum_I},~\eqref{eq:power},~\text{and} ~\eqref{eq:semi}.
\end{align}
\end{subequations}
% By introducing an auxiliary variable $u$, (P3.1) is rewritten as
% \begin{subequations}
% \begin{align} \nonumber
%   &\text{(P3.2)}: \max_{\{\mathbf w_k\}, \mathbf R_0,u} \quad u\\  
%    &\text{s.t.} \quad  \mathrm{tr}\left(\dot {\mathbf B} \left(\sum_{k\in\mathcal{K}} \mathbf w_k\mathbf w_k^{H} + \mathbf R_0\right) \dot {\mathbf B}^{H}\right)-\frac{\left|\mathrm{tr}\left(\mathbf B \left(\sum_{k\in\mathcal{K}} \mathbf w_k\mathbf w_k^{H} + \mathbf R_0\right) \dot {\mathbf B}^{H}\right)\right|^2}{\mathrm{tr}\left(\mathbf B\left(\sum_{k\in\mathcal{K}} \mathbf w_k\mathbf w_k^{H} + \mathbf R_0\right) \mathbf B^{H}\right)}>= u\\  \nonumber
%    & \qquad \eqref{eq:SINR_minimum_I},~\eqref{eq:power},~\text{and} ~\eqref{eq:semi}.
% \end{align}
% \end{subequations}
By introducing an auxiliary variable $u$ and using the Schur's complement, problem (P3.1) is reformulated as
\begin{subequations}
\begin{align} \nonumber
  &\text{(P3.2)}: \max_{\{\mathbf w_k\}, \mathbf R_0, u}  \quad  u  \\  
   & \text{s.t.} \quad \begin{bmatrix}
\mathrm{tr}\left(\dot {\mathbf B}\left(\sum_{k\in\mathcal{K}} \mathbf w_k\mathbf w_k^{H} + \mathbf R_0\right) \dot {\mathbf B}^{H}\right)-u & \mathrm{tr}\left(\mathbf B \left(\sum_{k\in\mathcal{K}} \mathbf w_k\mathbf w_k^{H} + \mathbf R_0\right) \dot {\mathbf B}^{H}\right)  \\
      \mathrm{tr}\left(\dot {\mathbf B}\left(\sum_{k\in\mathcal{K}} \mathbf w_k\mathbf w_k^{H} + \mathbf R_0\right) \mathbf B^{H}\right)& \mathrm{tr}\left(\mathbf B\left(\sum_{k\in\mathcal{K}} \mathbf w_k\mathbf w_k^{H} + \mathbf R_0\right) \mathbf B^{H}\right)
\end{bmatrix} \succeq \mathbf{0} \tag{33}\\\notag
   &\qquad ~ \eqref{eq:SINR_minimum_I},~\eqref{eq:power},~\text{and} ~\eqref{eq:semi}.
\end{align}
\end{subequations}

Next, we use the SDR technique to obtain the {\it optimal} solution to problem (P3.2). We first define $\mathbf  W_k = \mathbf  w_k \mathbf  w_k^{H}$ with $\mathbf  W_k \succeq \mathbf{0}$ and $\mathrm{rank}(\mathbf  W_k) \le 1$, and also define $\mathbf  H_k = \mathbf  h_{k}\mathbf  h_{k}^{H}, \forall k \in \mathcal{K}$. Then, problem (P3.2) is equivalently reformulated as 
\setcounter{equation}{33}
\begin{subequations}
\begin{align} \nonumber
  \text{(P3.3)}:&\max_{\{\mathbf W_k\}, \mathbf R_0, u}  \quad  u  \\  
   \text{s.t.} \label{eq:W_CRB_point}& \quad \begin{bmatrix}
\mathrm{tr}\left(\dot {\mathbf B}\left(\sum_{k\in\mathcal{K}} \mathbf W_k + \mathbf R_0\right) \dot {\mathbf B}^{H}\right)-u & \mathrm{tr}\left(\mathbf B \left(\sum_{k\in\mathcal{K}} \mathbf W_k + \mathbf R_0\right) \dot {\mathbf B}^{H}\right)  \\
      \mathrm{tr}\left(\dot {\mathbf B}\left(\sum_{k\in\mathcal{K}} \mathbf W_k+ \mathbf R_0\right) \mathbf B^{H}\right)& \mathrm{tr}\left(\mathbf B\left(\sum_{k\in\mathcal{K}} \mathbf W_k + \mathbf R_0\right) \mathbf B^{H}\right)
\end{bmatrix} \succeq \mathbf{0}\\
   &\quad (1+\frac{1}{\Gamma_k})\mathrm{tr}(\mathbf  H_k\mathbf  W_k)-\mathrm{tr}\left(\mathbf  H_k \left(\sum_{k\in\mathcal{K}}\mathbf  W_k+\mathbf  R_0\right)\right) \ge \sigma^2_k, \forall k \in \mathcal{K}\\ 
  \label{eq:W_energy_point}&\quad \sum_{k\in\mathcal{K}}\mathrm {tr}(\mathbf W_k) + \mathrm {tr}(\mathbf R_0) \le P_0 \\
     \label{eq:W_semi_point}&  \quad \mathbf R_0 \succeq \mathbf{0}, \mathbf  W_k \succeq \mathbf{0}, \forall k \in \mathcal{K}\\
  \label{eq:rank-one_W_point}& \quad \mathrm{rank}(\mathbf W_k) \le 1, \forall k \in \mathcal{K}.
\end{align}
\end{subequations}

Then, we drop the rank-one constraints in \eqref{eq:rank-one_W_point} to get the SDR version of (P3.3), denoted by (SDR3.3), which is a convex SDP and thus can be optimally solved by CVX. Let $\{\mathbf W^\star_k\}$ and  $\mathbf R^\star_0$ denote the optimal solution to (SDR3.3). We then have the following proposition. 
\begin{proposition} \label{prop:SDR_point}
The SDR of (P3.3) or equivalently (P3.1) is tight, i.e., problems (P3.1), (P3.2), (P3.3), and (SDR3.3) have the same optimal value.  Given the optimal solution $\{\mathbf W_k^{\star}\}$ and $\mathbf R_0^{\star}$ to (SDR3.3), the optimal solution to (P3.1) is
\begin{equation} \label{eq:w_new_point}
\mathbf w_k^\text{opt,I} = (\mathbf h_k^{H} \mathbf W_k^{\star} \mathbf h_k)^{-1/2}\mathbf W_k^{\star} \mathbf h_k, \forall k \in\mathcal{K},
\end{equation}
 \begin{equation} \label{eq:R_new_point}
\mathbf R_0^\text{opt,I} = \mathbf R_0^{\star} + \sum_{k\in \mathcal{K}} \mathbf W_k^{\star}- \sum_{k\in \mathcal{K}}\mathbf w_k^\text{opt,I}(\mathbf w_k^\text{opt,I})^H.
\end{equation}
\end{proposition}
\begin{IEEEproof}
The proof is similar to that of Proposition \ref{prop:SDR} and thus is omitted for brevity.
\end{IEEEproof}

Furthermore, we have the following proposition, which shows the solution structure of (SDR3.3).
\begin{proposition}\label{proposition_zero_point}
There exists one optimal solution to problem (SDR3.3), such that $\mathbf R_0^\star = \mathbf 0$.
\end{proposition}
\begin{IEEEproof}
The proof is similar to that of Proposition \ref{proposition_zero} and thus is omitted for brevity.
\end{IEEEproof}
However, the property in Proposition~\ref{proposition_zero_point} may not hold for problem (P3.3), (P3.2), or (P3.1), i.e., there may not exist any optimal solution for them with $\mathbf R_0^\text{opt,I}=\mathbf 0$ based on \eqref{eq:R_new_point}.

\subsubsection{Reflective Beamforming Optimization for (P3)}

Then, we optimize the reflective beamformer $\mathbf v$ in (P3) under any given transmit beamformers $\mathbf R_0$ and $\{\mathbf w_k\}$, in which the CRB expression in \eqref{eq:CRB_point_2} is used. By letting $\tilde{\mathbf G}_{k} = \mathrm {diag}(\mathbf{h}_{\text{r},k}^{H}) \mathbf G$, we have $\mathbf{h}_{\text{r},k}^{H} \mathbf{\Phi} \mathbf{G} =\mathbf{v}^{T}\tilde{\mathbf G}_{k}, \forall k \in \mathcal{K}$. The SINR in \eqref{eq:SINR} is thus rewritten as
\begin{equation}
    \gamma_k^{\text{I}}\!=\!\frac{(\mathbf{v}^{T}\tilde{\mathbf G}_{k}+\mathbf{h}_{\text{d},k}^{H})\mathbf{W}_k(\tilde{\mathbf G}_{k}^{H}\mathbf{v}^*+\mathbf{h}_{\text{d},k})}{\sum_{i\in\mathcal{K},i\neq k}(\mathbf{v}^{T}\tilde{\mathbf G}_{k}+\mathbf{h}_{\text{d},k}^{H})\mathbf W_i(\tilde{\mathbf G}_{k}^{H}\mathbf{v}^*+\mathbf{h}_{\text{d},k})\!+\!(\mathbf{v}^{T}\tilde{\mathbf G}_{k}+\mathbf{h}_{\text{d},k}^{H})\mathbf R_0(\tilde{\mathbf G}_{k}^{H}\mathbf{v}^*+\mathbf{h}_{\text{d},k}) + \sigma_k^2}.
\end{equation}
Then, the SINR constraints in \eqref{eq:SINR_minimum_I} are equivalent to 
$\tilde{\mathbf v}^H\mathbf{Q}_{k}^{\text{I}}\tilde{\mathbf v} \geq \Gamma_k \sigma_k^2, \forall k \in \mathcal{K}$
with  
\begin{equation}   
\mathbf{Q}_{k}^{\text{I}}=\left[                 
  \begin{array}{c}   
    \tilde{\mathbf G}_{k}^* \\ 
    \mathbf{h}_{\text{d},k}^{T}
  \end{array}
\right]\left(\mathbf{W}_k^*-\Gamma\left(\sum_{i\in\mathcal{K},i\neq k}\mathbf{W}_i^*+ \mathbf R_0^*\right)\right)
\left[                 
  \begin{array}{cc}   
    \tilde{\mathbf G}_{k}^T & \mathbf{h}_{\text{d},k}^{*}
  \end{array}
\right] ~\text{and}~
\tilde{\mathbf v}=
\left[                 
  \begin{array}{c}   
    \mathbf{v}\\ 
    1
  \end{array}
\right].
\end{equation}
After that, the reflective beamforming optimization problem is formulated as
\begin{subequations}
\begin{align} \nonumber
  \text{(P3.4)}:\max_{\mathbf v} & \quad  \mathbf v^{H} \mathbf{R}_{2} \mathbf v \left(\mathbf v^{H} \mathbf D \mathbf{R}_{1}  \mathbf D\mathbf v-\frac{|\mathbf v^{H} \mathbf D \mathbf{R}_{1} \mathbf v|^2}{\mathbf v^{H} \mathbf{R}_{1}  \mathbf v}\right) + \mathbf v^{H} \mathbf{R}_{1} \mathbf v \left(\mathbf v^{H} \mathbf D \mathbf{R}_{2} \mathbf D \mathbf v-\frac{|\mathbf v^{H} \mathbf D \mathbf{R}_{2} \mathbf v|^2}{\mathbf v^{H} \mathbf{R}_{2} \mathbf v}\right)\\  
   \text{s.t.}& \quad \tilde{\mathbf v}^H\mathbf{Q}_{k}^{\text{I}}\tilde{\mathbf v} \geq \Gamma_k \sigma_k^2, \forall k \in \mathcal{K}\\
   \label{eq:phase_v}& \quad |\mathbf v_n| =1 , \forall n \in \mathcal{N}.
\end{align}
\end{subequations}

% \begin{equation}
%     \mathrm{SINR}_k^{\text{II}}=\frac{(\mathbf{v}^{T}\tilde{\mathbf G}_{k}+\mathbf{h}_{\text{d},k}^{H})\mathbf{W}_k(\tilde{\mathbf G}_{k}^{H}\mathbf{v}^*+\mathbf{h}_{\text{d},k})}{\sum_{i\in\mathcal{K},i\neq k}(\mathbf{v}^{T}\tilde{\mathbf G}_{k}+\mathbf{h}_{\text{d},k}^{H})\mathbf W_i(\tilde{\mathbf G}_{k}^{H}\mathbf{v}^*+\mathbf{h}_{\text{d},k}) + \sigma^2} \ge  \Gamma_k,
% \end{equation}

% Then problem (P3.5) is equivalently re-expressed as
% \begin{subequations}
% \begin{align} \nonumber
%   \text{(P3.6)}:\max_{\mathbf v} \ & \  \mathbf v^{H} \mathbf{R}_{2}(\theta)  \mathbf v \left(\mathbf v^{H} \mathbf D \mathbf{R}_{1}(\theta)  \mathbf D\mathbf v-\frac{|\mathbf v^{H} \mathbf D \mathbf{R}_{1}(\theta) \mathbf v|^2}{\mathbf v^{H} \mathbf{R}_{1}(\theta)  \mathbf v}\right)\\ \notag
%   & \ + \mathbf v^{H} \mathbf{R}_{1}(\theta)  \mathbf v \left(\mathbf v^{H} \mathbf D \mathbf{R}_{2}(\theta) \mathbf D \mathbf v-\frac{|\mathbf v^{H} \mathbf D \mathbf{R}_{2}(\theta) \mathbf v|^2}{\mathbf v^{H} \mathbf{R}_{2}(\theta) \mathbf v}\right)\\  
%    \text{s.t.}& \quad {\tilde{\mathbf v}}^{{H}}\mathbf{Q}_{k}^{\text{I}}{\tilde{\mathbf v}} \geq \Gamma \sigma^2, \forall k \in \mathcal{K}\\
%    &\quad \tilde{\mathbf v} = [\mathbf v^T,1]^T\\
%       & \quad |\mathbf v_n|=1, \forall n\in \{1,\cdots,N\}.
% \end{align}
% \end{subequations}

% Next, we use  To
% resolve this issue, we first deal with constraint (29d) based on
% SDR, and then use SCA to approximate the relaxed problem.
Problem (P3.4) is non-convex due to the non-concavity of the objective function and the unit-modulus constraint in \eqref{eq:phase_v}. To resolve this issue, we first deal with the unit-modulus constraint \eqref{eq:phase_v} based on SDR, and then use the SCA to approximate the relaxed problem. We define $\tilde {\mathbf V}=\tilde{\mathbf v}{\tilde{\mathbf v}}^{{H}}$ with $\tilde{\mathbf V} \succeq \mathbf{0}$ and $\mathrm {rank}(\tilde{\mathbf V})=1$, 
\begin{equation} 
\tilde{\mathbf{R}}_{1}=
\left[                 
  \begin{array}{cc}   
    \mathbf{R}_{1}& \mathbf{0}_{N\times 1}\\ 
    \mathbf{0}_{1\times N}& 0\\ 
  \end{array}
\right],
\tilde{\mathbf R}_2=
\left[                 
  \begin{array}{cc}   
    \mathbf{R}_{2}& \mathbf{0}_{N\times 1}\\ 
    \mathbf{0}_{1\times N}& 0\\ 
  \end{array}
\right], ~{\text{and}}~
\tilde{\mathbf{D}}=
\left[                 
  \begin{array}{cc}   
    \mathbf{D}& \mathbf{0}_{N\times 1}\\ 
    \mathbf{0}_{1\times N}& 0\\ 
  \end{array}
\right].
\end{equation}
After that, problem (P3.4) is equivalently reformulated as
\begin{subequations}
\begin{align} \nonumber
  \text{(P3.5)}:\max_{\tilde{\mathbf V}} \  &  
    \quad \mathrm{tr}(\tilde{\mathbf R}_2\tilde{\mathbf V}) \mathrm{tr}( \tilde{\mathbf D} \tilde{\mathbf R}_1 \tilde{\mathbf D}\tilde{\mathbf V})-\mathrm{tr}(\tilde{\mathbf R}_2\tilde{\mathbf V})\frac{|\mathrm{tr}(\tilde{\mathbf D} \tilde{\mathbf R}_1\tilde{\mathbf V})|^2}{\mathrm{tr}(\tilde{\mathbf R}_1\tilde{\mathbf V})}   \\ \notag
   &\quad +\mathrm{tr}(\tilde{\mathbf R}_1\tilde{\mathbf V}) \mathrm{tr}(\tilde{\mathbf D} \tilde{\mathbf R}_2 \tilde{\mathbf D}\tilde{\mathbf V})-\mathrm{tr}(\tilde{\mathbf R}_1\tilde{\mathbf V}) \frac{|\mathrm{tr}(\tilde{\mathbf D} \tilde{\mathbf R}_2\tilde{\mathbf V})|^2}{\mathrm{tr}(\tilde{\mathbf R}_2\tilde{\mathbf V})} \\\label{eq:SINR-constrained_V}
   \text{s.t.}&\quad \mathrm{tr}\left(\mathbf{Q}_{k}^{\text{I}}\tilde{\mathbf V}\right) \geq \Gamma_k \sigma_k^2,\forall k \in \mathcal{K}\\\label{eq:1_constraint_V}
   &\quad \tilde{\mathbf V}_{n,n}=1, \forall n\in \{1,\cdots,N+1\}\\\label{eq:semi_constraint_V}
    & \quad \tilde{\mathbf V} \succeq \mathbf{0}\\
    \label{eq:rank-one-V}& \quad \mathrm {rank}(\tilde{\mathbf V})=1.
\end{align}
\end{subequations}
By relaxing the rank-one constraint in \eqref{eq:rank-one-V}, the SDR version of problem (P3.5) is obtained as (SDR3.5).
% \begin{subequations}
% \begin{align} \nonumber
%   \text{(SDR3.6)}:\max_{\tilde{\mathbf V}}& \quad \mathrm{tr}(\tilde{\mathbf R}_2\tilde{\mathbf V}) \mathrm{tr}( \tilde{\mathbf D} \tilde{\mathbf R}_1 \tilde{\mathbf D}\mathbf V)-\mathrm{tr}(\tilde{\mathbf R}_2\tilde{\mathbf V})\frac{|\mathrm{tr}(\tilde{\mathbf D} \tilde{\mathbf R}_1\tilde{\mathbf V})|^2}{\mathrm{tr}(\tilde{\mathbf R}_1\tilde{\mathbf V})} \\  \notag
%   &\quad +\mathrm{tr}(\tilde{\mathbf R}_1\tilde{\mathbf V}) \mathrm{tr}(\tilde{\mathbf D} \tilde{\mathbf R}_2 \tilde{\mathbf D}\tilde{\mathbf V})-\mathrm{tr}(\tilde{\mathbf R}_1\tilde{\mathbf V}) \frac{|\mathrm{tr}(\tilde{\mathbf D} \tilde{\mathbf R}_2\tilde{\mathbf V})|^2}{\mathrm{tr}(\tilde{\mathbf R}_2\tilde{\mathbf V})}\\\notag
%     \text{s.t.}&\quad \eqref{eq:SINR-constrained_V},~\eqref{eq:1_constraint_V},~\text{and}~\eqref{eq:semi_constraint_V}.
% \end{align}
% \end{subequations}
Then, by introducing two auxiliary variables $t_{1}$ and $t_{2}$ and using the Schur's complement, problem (SDR3.5) is equivalently re-expressed as
\begin{subequations}
\begin{align} \nonumber
  \text{(SDR3.6)}:\max_{\tilde{\mathbf V}, t_{1},t_{2}}& \quad \mathrm{tr}(\tilde{\mathbf R}_2\tilde{\mathbf V}) \mathrm{tr}( \tilde{\mathbf D} \tilde{\mathbf R}_1\tilde{\mathbf D}\mathbf V)-\mathrm{tr}(\tilde{\mathbf R}_2\tilde{\mathbf V})t_{1} +\mathrm{tr}(\tilde{\mathbf R}_1\tilde{\mathbf V}) \mathrm{tr}(\tilde{\mathbf D} \tilde{\mathbf R}_2 \tilde{\mathbf D}\tilde{\mathbf V})-\mathrm{tr}(\tilde{\mathbf R}_1\tilde{\mathbf V}) t_{2}\\  \label{eq:Semi_schur_I}
    \text{s.t.}&\quad \left[\begin{array}{cc}
      t_{1} & \mathrm{tr}(\tilde{\mathbf D} \tilde{\mathbf R}_1\tilde{\mathbf V})  \\ 
      \mathrm{tr}(\tilde{\mathbf V}^{H}\tilde{\mathbf R}_1^H\tilde{\mathbf D}^H ) &  \mathrm{tr}(\tilde{\mathbf R}_1\tilde{\mathbf V})
\end{array}\right] \succeq \mathbf{0}\\ \label{eq:Semi_schur_II}
   &\quad \left[\begin{array}{cc}
      t_{1} & \mathrm{tr}(\tilde{\mathbf D} \tilde{\mathbf R}_2\tilde{\mathbf V})  \\
      \mathrm{tr}(\tilde{\mathbf V}^{H}\tilde{\mathbf R}_2^H\tilde{\mathbf D}^H ) &  \mathrm{tr}(\tilde{\mathbf R}_2\tilde{\mathbf V})
\end{array}\right] \succeq \mathbf{0}\\ \notag
   &\quad \eqref{eq:SINR-constrained_V},~\eqref{eq:1_constraint_V},~\text{and}~\eqref{eq:semi_constraint_V}.
\end{align}
\end{subequations}
To deal with the non-concave objective function in (SDR3.6), we first reformulate the objective function as $f_{1}(\tilde{\mathbf V},t_{1},t_{2})+f_{2}(\tilde{\mathbf V},t_{1},t_{2})$
% \begin{equation}
% \begin{split}
% &\mathrm{tr}(\tilde{\mathbf R}_2(\theta)\tilde{\mathbf V}) \mathrm{tr}( \tilde{\mathbf D} \tilde{\mathbf R}_1(\theta) \tilde{\mathbf D}\mathbf V)-\mathrm{tr}(\tilde{\mathbf R}_2(\theta)\tilde{\mathbf V})t_{1}  +\mathrm{tr}(\tilde{\mathbf R}_1(\theta)\tilde{\mathbf V}) \mathrm{tr}(\tilde{\mathbf D} \tilde{\mathbf R}_2(\theta) \tilde{\mathbf D}\tilde{\mathbf V})-\mathrm{tr}(\tilde{\mathbf R}_1(\theta)\tilde{\mathbf V}) t_{2}\\
% =&f_{1}(\mathbf V,t_{1},t_{2})+f_{2}(\mathbf V,t_{1},t_{2})
% \end{split}
% \end{equation}
with 
\begin{equation}
\begin{split}
f_{1}(\tilde{\mathbf V},t_{1},t_{2})
=&\frac{1}{4}\left(\mathrm{tr}((\tilde{\mathbf R}_2+\tilde{\mathbf D}\tilde{\mathbf R}_1\tilde{\mathbf D})\tilde{\mathbf V})\right)^2+\frac{1}{4}\left(\mathrm{tr}(\tilde{\mathbf R}_2\tilde{\mathbf V})-t_{1}\right)^2\\
  &+\frac{1}{4}\left(\mathrm{tr}((\tilde{\mathbf R}_1+\tilde{\mathbf D}\tilde{\mathbf R}_2\tilde{\mathbf D})\tilde{\mathbf V})\right)^2+\frac{1}{4}\left(\mathrm{tr}(\tilde{\mathbf R}_1 \tilde{\mathbf V})-t_{2}\right)^2,
\end{split}
\end{equation}
\begin{equation}
\begin{split}
f_{2}(\tilde{\mathbf V},t_{1},t_{2})
=&-\frac{1}{4}\left(\mathrm{tr}((\tilde{\mathbf R}_2-\tilde{\mathbf D}\tilde{\mathbf R}_1\tilde{\mathbf D})\tilde{\mathbf V})\right)^2-\frac{1}{4}\left(\mathrm{tr}(\tilde{\mathbf R}_2\tilde{\mathbf V})+t_{1}\right)^2\\
  &-\frac{1}{4}\left(\mathrm{tr}((\tilde{\mathbf R}_1-\tilde{\mathbf D}\tilde{\mathbf R}_2\tilde{\mathbf D})\tilde{\mathbf V})\right)^2-\frac{1}{4}\left(\mathrm{tr}(\tilde{\mathbf R}_1 \tilde{\mathbf V})+t_{2}\right)^2.
\end{split}
\end{equation}
% \begin{equation}
% \begin{split}
% f_{2}(\tilde{\mathbf V},t_{1},t_{2})
% =&-\frac{1}{4}\mathrm{tr}^2((\tilde{\mathbf R}_2-\tilde{\mathbf D}\tilde{\mathbf R}_1\tilde{\mathbf D})\tilde{\mathbf V})-\frac{1}{4}(\mathrm{tr}(\tilde{\mathbf R}_2 \tilde{\mathbf V})+t_{1})^2\\
%   &-\frac{1}{4}\mathrm{tr}^2((\tilde{\mathbf R}_1-\tilde{\mathbf D}\tilde{\mathbf R}_2\tilde{\mathbf D})\tilde{\mathbf V})-\frac{1}{4}(\mathrm{tr}(\tilde{\mathbf R}_1 \tilde{\mathbf V})+t_{2})^2.
% \end{split}
% \end{equation}
Here in, $f_{1}(\tilde{\mathbf V},t_{1},t_{2})$ and  $f_{2}(\tilde{\mathbf V},t_{1},t_{2})$ are convex and concave functions, respectively. Then, we use SCA to approximate the non-concave function $f_{1}(\tilde{\mathbf V},t_{1},t_{2})$ in an iterative manner. In each inner iteration $r$, with local point $\mathbf V^{(r)},t_1^{(r)},t_2^{(r)}$ , we obtain a global linear lower bound function $\hat f_{1}^{(r)}(\tilde{\mathbf V},t_{1},t_{2})$ for $f_{1}(\tilde{\mathbf V},t_{1},t_{2})$ using its first-order Taylor expansion, i.e.,
\begin{equation}
\begin{split}
f_1(\tilde{\mathbf V},t_1,t_2)\ge& f_1(\tilde{\mathbf V}^{(r)},t_1^{(r)},t_2^{(r)})+\frac{1}{2}\mathrm{tr}((\tilde{\mathbf R}_2+\tilde{\mathbf D}\tilde{\mathbf R}_1\tilde{\mathbf D})\tilde{\mathbf V}^{(r)})\mathrm{tr}((\tilde{\mathbf R}_2\!+\!\tilde{\mathbf D}\tilde{\mathbf R}_1\tilde{\mathbf D})(\tilde{\mathbf V}-\tilde{\mathbf V}^{(r)}))\\
&+\frac{1}{2}\mathrm{tr}((\tilde{\mathbf R}_1+\tilde{\mathbf D}\tilde{\mathbf R}_2\tilde{\mathbf D})\tilde{\mathbf V}^{(r)})\mathrm{tr}((\tilde{\mathbf R}_1+\tilde{\mathbf D}\tilde{\mathbf R}_2\tilde{\mathbf D})(\tilde{\mathbf V}-\tilde{\mathbf V}^{(r)}))\\
&+\frac{1}{2}t_1^{(r)}(t_1-t_1^{(r)})+\frac{1}{2}t_2^{(r)}(t_2-t_2^{(r)})+\frac{1}{2}\mathrm{tr}(\tilde{\mathbf R}_2\tilde{\mathbf V}^{(r)})\mathrm{tr}(\tilde{\mathbf R}_2(\tilde{\mathbf V}-\tilde{\mathbf V}^{(r)}))\\
&+\frac{1}{2}\mathrm{tr}(\tilde{\mathbf R}_1\tilde{\mathbf V}^{(r)})\mathrm{tr}(\tilde{\mathbf R}_1(\tilde{\mathbf V}-\tilde{\mathbf V}^{(r)}))-\frac{1}{2}\mathrm{tr}(\tilde{\mathbf R}_2(\tilde{\mathbf V}-\tilde{\mathbf V}^{(r)}))t_1^{(r)}\\
&-\frac{1}{2}\mathrm{tr}(\tilde{\mathbf R}_1(\tilde{\mathbf V}-\tilde{\mathbf V}^{(r)}))t_2^{(r)}-\frac{1}{2}\mathrm{tr}(\tilde{\mathbf R}_2\tilde{\mathbf V}^{(r)})(t_1-t_1^{(r)})-\frac{1}{2}\mathrm{tr}(\tilde{\mathbf R}_1\tilde{\mathbf V}^{(r)})(t_2-t_2^{(r)})\\
\triangleq &\hat f_1^{(r)}(\tilde{\mathbf V},t_1,t_2).
\end{split}
\end{equation}

Replacing $f_{1}(\tilde{\mathbf V},t_{1},t_{2})$ by $\hat f_{1}^{(r)}(\tilde{\mathbf V},t_{1},t_{2})$, problem (SDR3.6) is approximated as the following convex form in inner iteration $r$:
\begin{subequations}
\begin{align} \nonumber
  (\text{SDR3.6}.r):\max_{\tilde{\mathbf V},t_{1},t_{2}}& \quad \hat f_{1}^{(r)}(\tilde{\mathbf V},t_{1},t_{2}) +   f_{2}(\tilde{\mathbf V},t_{1},t_{2}) \\  \notag
    \text{s.t.}&\quad  \eqref{eq:Semi_schur_I},~\eqref{eq:Semi_schur_II},~ \eqref{eq:SINR-constrained_V},~\eqref{eq:1_constraint_V},~\text{and}~\eqref{eq:semi_constraint_V}.
\end{align}
\end{subequations}
Problem $(\text{SDR3.6}.r)$ is convex and thus can be optimally solved by CVX. Let $\tilde{\mathbf V}^{(r,\star)}$, $t_1^{(r,\star)}$, and $t_2^{(r,\star)}$ denote the optimal solution to problem (SDR3.6.$r$), which is then updated to be the local point $\tilde{\mathbf V}^{(r+1)}$, $t_1^{(r+1)}$, and $t_2^{(r+1)}$ for the next inner iteration $r+1$. Since $\hat f_1^{(r)}(\tilde{\mathbf V},t_1,t_2)$ serves as a lower bound of $f_1(\tilde{\mathbf V},t_1,t_2)$, we have
$f_1(\tilde{\mathbf V}^{(r+1)}, t_1^{(r+1)}, t_2^{(r+1)}) + f_2(\tilde{\mathbf V}^{(r+1)}, t_1^{(r+1)}, t_2^{(r+1)})
\ge  \hat f_{1}^{(r)}(\tilde{\mathbf V}^{(r)}, t_1^{(r)}, t_2^{(r)})+ f_2(\tilde{\mathbf V}^{(r)}, t_1^{(r)}, t_2^{(r)})
= f_1(\tilde{\mathbf V}^{(r)}, t_1^{(r)}, t_2^{(r)})+ f_2(\tilde{\mathbf V}^{(r)}, t_1^{(r)}, t_2^{(r)})$.
Thus, the inner iteration leads to a non-decreasing objective value for problem (SDR3.6).
Therefore, the convergence of SCA for solving problem (SDR3.6) is ensured. Let $\tilde{\mathbf V}^\star$, $t_1^\star$, and $t_2^\star$ denote the obtained converged  solution to problem (SDR3.6) using SCA, where $\mathrm{rank}(\tilde{\mathbf V}^\star) >1$ holds in general. 

Finally, Gaussian randomization is used to construct an approximate rank-one solution of $\tilde{\mathbf V}$ to problem (P3.5). Motivated by that in\cite{xianxin}, we first generate a number of randomizations $\mathbf r \sim \mathcal{CN}(\mathbf{0},\tilde{\mathbf V}^\star)$, and accordingly construct a series of candidate solutions as $\mathbf{v}=e^{j\mathrm {arg}([\frac{\mathbf r}{\mathbf r_{N+1}}]_{(1:N)})}$. By independently generating Gaussian random vector $\mathbf{r}$ multiple times, we obtain the solution of (P3.5) as the one achieving the maximum objective value of (P3.5) while satisfying the SINR constraints among all these random realizations.

\begin{table}
\centering{
\caption{Proposed Algorithm for Solving Problem (P3)\label{tab:table2}}}
\vspace{-0.5cm}
 \hrule
\vspace{0.1cm} \textbf{Algorithm 2}  \vspace{0.1cm}
\hrule 
\begin{enumerate}[a)]
            \item Set outer iteration index $l= 1$ and initialize the reflective beamformer with random phase shifts as $\mathbf{v}^{(l)}$. 
            \item \textbf{Repeat}: \begin{enumerate}[1)]
            				\item Under given reflective beamformer $\mathbf{v}^{(l)}$, solve problem (SDR3.3) to obtain the optimal solution as $\mathbf R_0^{\star(l)}$, $\{\mathbf w_k^{\star(l)}\}$. 
            				\item Construct the optimal rank-one solution $\{\mathbf w_k^{(l)}\}$ and $ \mathbf R_0^{(l)}$ to (P3.1) based on $\{{\mathbf W}_k^{\star(l)}\}$ and $ {\mathbf R}_0^{\star(l)}$ by using Proposition~\ref{prop:SDR_point}.
            				\item Set inner iteration index  $r=1$, $\tilde{\mathbf V}^{(r)}=\tilde{\mathbf v}^{(l)}(\tilde{\mathbf v}^{(l)})^{H}$, $t_1^{(r)}= \frac{|\mathrm{tr}(\tilde{\mathbf D} \tilde{\mathbf R}_1(\theta)\tilde{\mathbf V})|^2}{\mathrm{tr}(\tilde{\mathbf R}_1(\theta)\tilde{\mathbf V})} $, and $t_2^{(r)}= \frac{|\mathrm{tr}(\tilde{\mathbf D} \tilde{\mathbf R}_2(\theta)\tilde{\mathbf V})|^2}{\mathrm{tr}(\tilde{\mathbf R}_2(\theta)\tilde{\mathbf V})} $.
            				\item \textbf{Repeat}: \begin{enumerate}[i)]
            						\item Construct function $\hat f_1^{(r)}(\tilde{\mathbf V},t_1,t_2)$ using $\tilde{\mathbf V}^{(r)}$, $t_1^{(r)}$, and $t_2^{(r)}$.
            						\item Solve problem (P3.6.$r$)  under given $\mathbf R_0^{(l)}$, $\{\mathbf w_k^{(l)}\}$ to obtain the optimal solution as $\tilde{\mathbf V}^{(r+1)}$, $t_1^{(r+1)}$, and $t_2^{(r+1)}$.
            						\item Update $r=r+1$.
            				\end{enumerate}
            				\item \textbf{Until} the convergence criterion is met or the maximum number of inner iterations is reached.
            				\item Construct an approximate rank-one solution $\mathbf{v}^{(l+1)}$ to problem (P3.5) via Gaussian randomization.
            				\item Update $l=l+1$.
            				\end{enumerate}
            				\item \textbf{Until} the convergence criterion is met or the maximum number of outer iterations is reached.
            			
        \end{enumerate}
\vspace{0.1cm} \hrule \vspace{-20pt}\label{algorithm:2}
\end{table}

\subsubsection{Complete Algorithm for Solving (P3)}
By combining the transmit and reflective beamforming designs in Sections IV-A-(1) and IV-A-(2), together with the alternating optimization, we have the complete algorithm to solve (P3), which is summarized as Algorithm~2 in Table~\ref{tab:table2}.
Notice that in each iteration of Algorithm~2, (P3.1) is optimally solved, which leads to a non-increasing CRB value. With sufficient number of Gaussian randomizations, the SDR approach achieves at least $\frac{\pi}{4}$-approximation of the optimal objective value of (P3.5) \cite{luo2010semidefinite}, thus leading to a monotonically non-increasing CRB value in general. Otherwise, the outer iteration is terminated. As a result, the convergence of Algorithm~2 for solving (P3) is ensured.

\subsection{Proposed Solution to Problem (P4) with Type-II CU Receivers}
In this subsection, we consider problem (P4) with Type-II CU receivers, which can be solved similarly as Algorithm 2 for (P3) based on  alternating optimization. Therefore, we present the transmit and reflective beamforming solution in the following briefly.

First, we optimize the transmit beamformers $\{\mathbf w_k\}$ and $\mathbf R_0$ in (P4) with any given reflective beamformer $\mathbf v$, which is given by 
\begin{subequations}
\begin{align} \nonumber
  \text{(P4.1)}: \max_{\{\mathbf w_k\}, \mathbf R_0} & \quad  \mathrm{tr}\left(\dot {\mathbf B} \left(\sum_{k\in\mathcal{K}} \mathbf w_k\mathbf w_k^{H} + \mathbf R_0\right) \dot {\mathbf B}^{H}\right)-\frac{\left|\mathrm{tr}\left(\mathbf B \left(\sum_{k\in\mathcal{K}} \mathbf w_k\mathbf w_k^{H} + \mathbf R_0\right) \dot {\mathbf B}^{H}\right)\right|^2}{\mathrm{tr}\left(\mathbf B \left(\sum_{k\in\mathcal{K}} \mathbf w_k\mathbf w_k^{H} + \mathbf R_0\right) \mathbf B^{H}\right)}\\ \notag  
   \text{s.t.}& \quad \eqref{eq:SINR_minimum_II},~\eqref{eq:power},~\text{and} ~\eqref{eq:semi}.
\end{align}
\end{subequations}
By introducing an auxiliary variable $u$, using the Schur's complement, and defining $\mathbf  W_k = \mathbf  w_k \mathbf  w_k^{H}$ with $\mathbf  W_k \succeq \mathbf{0}$ and $\mathrm{rank}(\mathbf  W_k) \le 1, \forall k\in \mathcal{K}$, (P4.1) is reformulated as
\begin{subequations}
\begin{align} \nonumber
  \text{(P4.2)}:&\max_{\{\mathbf W_k\}, \mathbf R_0, u}  \quad  u  \\  
   \text{s.t.}&\quad \frac{1}{\Gamma_k}\mathrm{tr}(\mathbf  H_k\mathbf  W_k)-\sum_{i\in\mathcal{K},i\neq k}\mathrm{tr}(\mathbf  H_k\mathbf  W_i) \ge \sigma^2_k, \forall k \in \mathcal{K} \tag{46}\\ \notag
  &\quad \eqref{eq:W_CRB_point},~\eqref{eq:W_energy_point},~\eqref{eq:W_semi_point},~\text{and}~ \eqref{eq:rank-one_W_point}.
\end{align}
\end{subequations}
We drop the rank-one constraints in \eqref{eq:rank-one_W_point} and accordingly express the SDR of (P4.2) as (SDR4.2), which is a convex SDP that can be optimally solved via CVX. Let $\{\mathbf W_k^{\star\star}\}$ and ${\mathbf R}_0^{\star\star}$ denote the optimal solution to (SDR4.2).  We then have the following proposition. 
\setcounter{equation}{46}
\begin{proposition} 
The SDR of (P4.2) or equivalently (P4.1) is tight, i.e., problems (P4.1), (P4.2), and (SDR4.2) have the same optimal value.  Given the optimal solution $\{\mathbf W_k^{\star\star}\}$ and ${\mathbf R}_0^{\star\star}$ to (SDR4.2), the optimal solution to (P4.1) is
\begin{equation} 
\mathbf w_k^\text{opt,II} = (\mathbf h_k^{H} \mathbf W_k^{\star\star} \mathbf h_k)^{-1/2}\mathbf W_k^{\star\star} \mathbf h_k, \forall k \in \mathcal{K},
\end{equation}
 \begin{equation}
\mathbf R_0^\text{opt,II} ={\mathbf R}_0^{\star\star} +  \sum_{k\in \mathcal{K}} \mathbf W_k^{\star\star}- \sum_{k\in \mathcal{K}}\mathbf w_k^\text{opt,II}(\mathbf w_k^\text{opt,II})^H.
\end{equation}
\end{proposition}
\begin{IEEEproof}
The proof is similar to that of Proposition~\ref{prop:SDR_2} and thus is omitted for brevity.
\end{IEEEproof}
It is worth noting that at the optimality, it generally holds that $\mathbf R_0^{\star\star} \neq \mathbf 0$ for problem (SDR4.2) and $\mathbf R_0^\text{opt,II} \neq \mathbf 0$ for problem (P4.1). This is different from problem (SDR3.3) and means that the dedicated sensing signals are generally needed when Type-II CU receivers are considered, similarly as for problem (SDR2.1).

Next, we optimize the reflective beamformer $\mathbf v$ in (P4) with any given transmit beamformers $\{\mathbf w_k\}$ and $\mathbf R_0$, for which the optimization problem becomes
\begin{subequations}
\begin{align} \nonumber
  \text{(P4.3)}:\max_{\mathbf v} & \quad  \mathbf v^{H} \mathbf{R}_{2} \mathbf v \left(\mathbf v^{H} \mathbf D \mathbf{R}_{1}  \mathbf D\mathbf v-\frac{|\mathbf v^{H} \mathbf D \mathbf{R}_{1} \mathbf v|^2}{\mathbf v^{H} \mathbf{R}_{1}  \mathbf v}\right) + \mathbf v^{H} \mathbf{R}_{1} \mathbf v \left(\mathbf v^{H} \mathbf D \mathbf{R}_{2} \mathbf D \mathbf v-\frac{|\mathbf v^{H} \mathbf D \mathbf{R}_{2} \mathbf v|^2}{\mathbf v^{H} \mathbf{R}_{2} \mathbf v}\right)\\  
   \text{s.t.}& \quad \tilde{\mathbf v}^H\mathbf{Q}_{k}^{\text{II}}\tilde{\mathbf v} \geq \Gamma_k \sigma_k^2, \forall k \in \mathcal{K}\\
   & \quad |\mathbf v_n| =1 , \forall n \in \mathcal{N},
\end{align}
\end{subequations}
where  
\begin{equation}   
\mathbf{Q}_{k}^{\text{II}}=\left[                 
  \begin{array}{c}   
    \tilde{\mathbf G}_{k}^* \\ 
    \mathbf{h}_{\text{d},k}^{T}
  \end{array}
\right]\left(\mathbf{W}_k^*-\Gamma\sum_{i\in\mathcal{K},i\neq k}\mathbf{W}_i^*\right)
\left[                 
  \begin{array}{cc}   
    \tilde{\mathbf G}_{k}^T & \mathbf{h}_{\text{d},k}^{*}
  \end{array}
\right] ~\text{and}~
\tilde{\mathbf v}=
\left[                 
  \begin{array}{c}   
    \mathbf{v}\\ 
    1
  \end{array}
\right].
\end{equation}
As problem (P4.3) has a similar structure as (P3.4), it can be solved similarly by using the algorithm in Section IV-A-(2), for which the details are omitted. Therefore, by combining the solutions to (P4.1) and (P4.3) together with alternating optimization, problem (P4) is finally solved.

% Next, consider the reflective beamforming optimization problem, which is same as (P3.5), denoted by (P4.3). This problem can be solved based on the design in Section IV-A2, for which the details are omitted for brevity. Therefore, by combining the solutions to (P4.1) and (P4.3) together with alternating optimization, problem (P4) is finally solved. 
\subsection{Solution Structure}

In this subsection, we analyze the solution structures of problems (P3) and (P4) to gain more insights. First, we consider the special case with one single CU. In this case, we have the following proposition, which is similar as Propositions \ref{prop:single_user_Type-I} and \ref{prop:single-user}, an thus its proof are omitted. 
\begin{proposition} \label{prop:single_user_Type-I_point}
At the optimality of (P3) with Type-I CU receivers, when $K=1$, the dedicated sensing beamformers should lie in the null space of the channel vector $\mathbf h_1$ from the BS to the CU, i.e., $\mathbf h_1^H \mathbf R_0^\text{opt,I} \mathbf h_1=0$. In this case, problems (P3) and (P4) become identical.
\end{proposition} 

Propositions \ref{prop:single_user_Type-I_point} shows that for the special case with one single CU, Type-II CU receivers do not provide any performance gain over their Type-I counterparts. 

Next, we discuss the solution structure of $\mathbf R_0$ for problems (P3) and (P4). It is observed from extensive simulations that when $K\ge 2$, $\mathbf R_0^\text{opt,I}= \mathbf 0$ occurs for problem (P3) but $\mathbf R_0^\text{opt,II} \neq \mathbf 0$ occurs for problem (P4). This shows that dedicated sensing signals are necessary only for Type-II CU receivers in the case of point target. This is consistent with the result with extended target in Section III.

\section{Numerical Results}
In this section, we provide numerical results to validate the performance of our proposed joint transmit and reflective beamforming design. We consider the distance-dependent path loss model, i.e.,
$L(d)=K_0\left(\frac{d}{d_0}\right)^{-\alpha_0}$,
where $K_0=-30$~dB is the pathloss at the reference distance $d_0 =1$~m and $\alpha_0$ is the path loss exponent. We set $\alpha_0$ as $2.2$, $2.2$, and $3.0$ for the BS-IRS, IRS-CU, and BS-CU links, respectively. We consider the Rician fading for the BS-IRS, IRS-CU, and BS-CU links with the Rician factor being $0.5$. Also, additional shadow fading is considered for the BS-CU links, with a standard deviation of $10~ \text{dB}$. The BS and the IRS are located at coordinate $(0~\text{m},0~\text{m})$ and $(4~\text{m},5~\text{m})$, respectively. The CUs are randomly located at  a rectangular grid with corners $(40~\text{m},0~\text{m})$, $(40~\text{m},-10~\text{m})$, $(50~\text{m},-10~\text{m})$, and $(50~\text{m},0~\text{m})$. For the point target case, the target is located at coordinate $(4~\text{m},1~\text{m})$ (i.e., the target DoA is $\theta=0$ w.r.t. the IRS), with a unit RCS. Without loss of generality, we assume that all CUs have the same SINR requirements, i.e., $\Gamma_k = \Gamma, \forall k \in \mathcal{K}$. We set $M=8$, $N=8$, $T=256$, $P_0 =30~\text{dBm}$, $\sigma_\text{R}^2 = -110~\text{dBm}$, and $\sigma_k^2 = -80~\text{dBm}, \forall k \in \mathcal{K}$. In the simulation, the results are obtained by averaging over $50$ independent realizations of the fading channel with different user locations.

% \begin{figure}[t]
%     \centering
%     \includegraphics[width=0.4\textwidth]{Fig/position.pdf}
%     \caption{Simulation setup.}
% 	\label{Simulation setup}
% \end{figure}

For performance comparison, we consider the following benchmark schemes.

\subsubsection{Transmit beamforming only with random IRS phase shifts (Transmit BF only)} This scheme optimizes the transmit beamformers $\{\mathbf w_k\}$ and $\mathbf R_0$ at the BS under given random phase shifts at IRS. 
%This method is implemented by using (SDR1.2) and Proposition 1 (or (SDR2.1) and Proposition 2) for the case with Type-I (II) CU receivers.

\subsubsection{Separate communication and sensing beamforming design (Separate BF design)} This scheme optimizes the transmit communication and sensing beamformers separately. First, we optimize the transmit communication beamformers $\{\bar{\mathbf w}_k\}$ and the reflective beamformer $\mathbf \Phi$ to minimize the transmit power while ensuring the SINR constraints at the CUs\cite{8811733}, and accordingly set $\mathbf w_k = \bar{\alpha} \bar{\mathbf w}_k, \forall k\in \mathcal{K}$, with $\bar{\alpha}\ge 1$ being an optimization variable to be decided. Then, we optimize the transmit sensing beamformer $\mathbf R_0$ together with $\bar{\alpha}$ to minimize the estimation CRB, subject to the minimum SINR constraints at individual CUs and the maximum transmit power constraint at the BS.

% \subsubsection{Joint beamforming design for minimum beampattern gain maximization (Beampattern design)} This scheme jointly optimizes the transmit beamformers $\{\mathbf w_k\}$ and $\mathbf R_0$ at the BS, and the the reflective beamformer $\mathbf v$ at the IRS to minimize the minimum beampattern gain at the desired sensing angles while ensuring the SINR constraints at the CUs\cite{song2021joint}.

\begin{figure}[t]
	\centering
	\begin{minipage}{0.45\linewidth}
		\centering
		\includegraphics[width=0.9\linewidth]{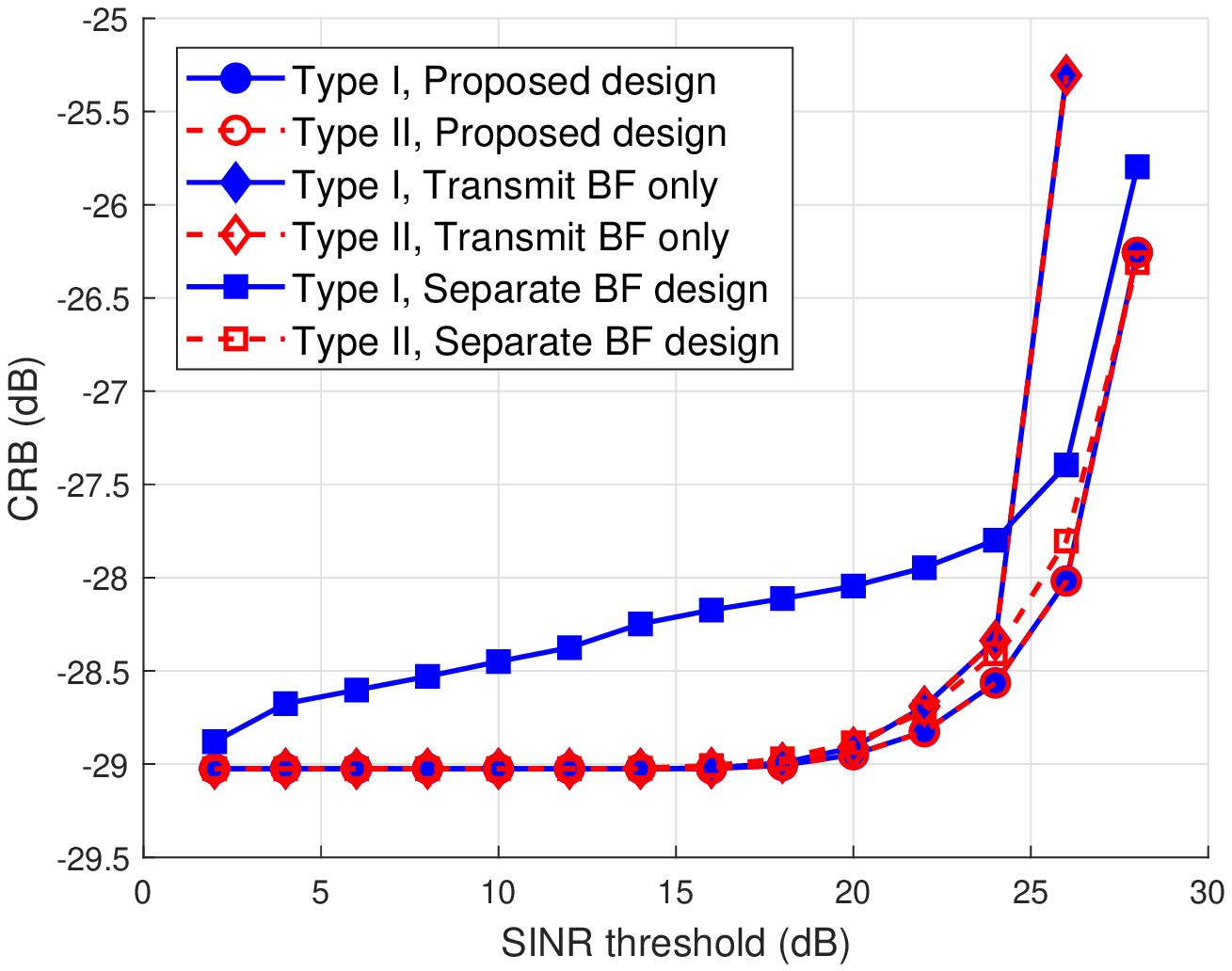}
		 \caption{The CRB for target response matrix estimation versus the SINR threshold $\Gamma$ with extended target, where $K=1$.}
    	\label{fig:CRB_SNR_single}
	\end{minipage}
	\quad
	\begin{minipage}{0.45\linewidth}
		\centering
		\includegraphics[width=0.9\linewidth]{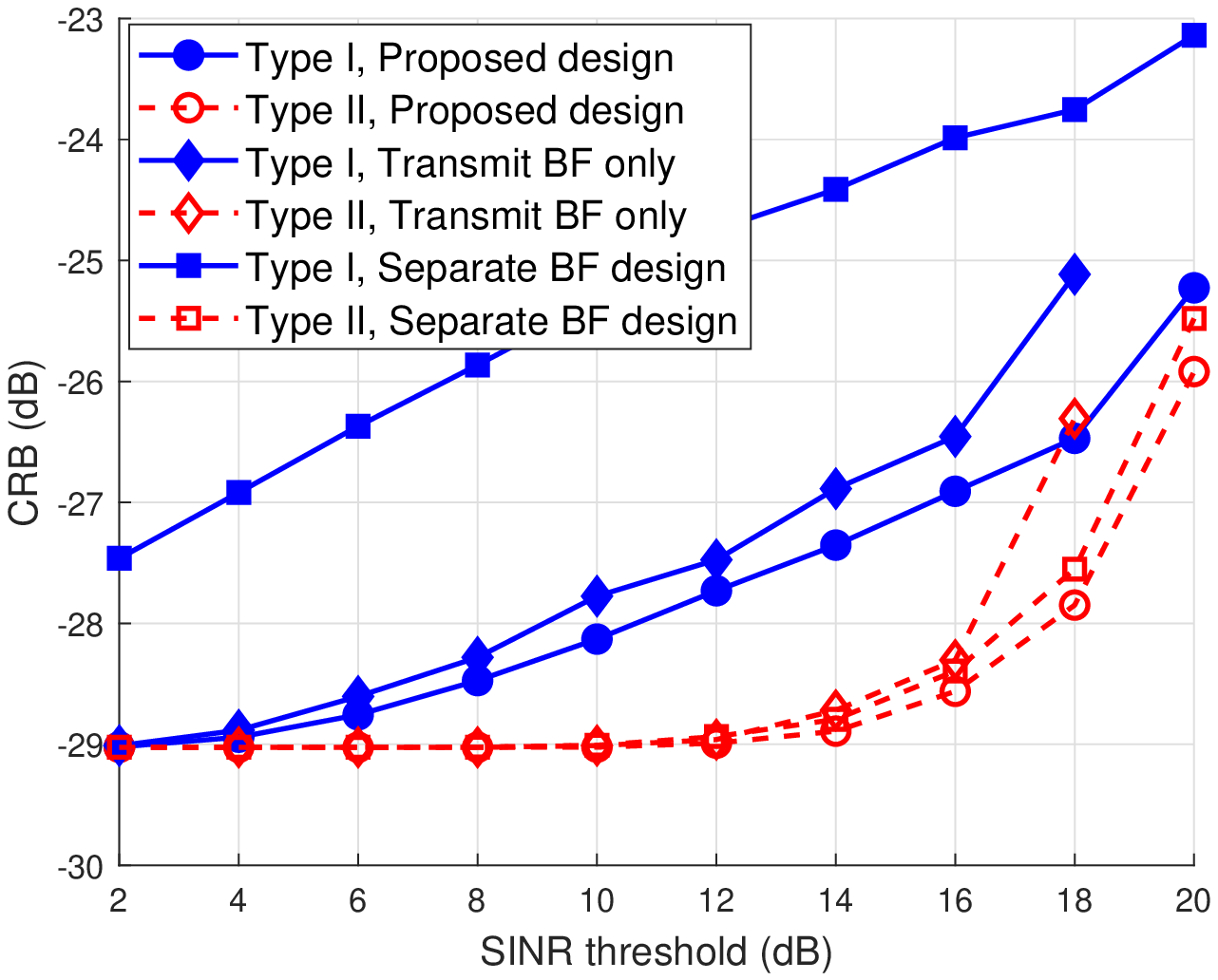}
		\caption{The CRB for target response matrix estimation versus the SINR threshold $\Gamma$ with extended target, where $K=4$.}
    	\label{fig:CRB_SNR_multi}
	\end{minipage}
\end{figure}
\begin{figure}[t]
	\centering
	\begin{minipage}{0.45\linewidth}
		\centering
		\includegraphics[width=0.9\linewidth]{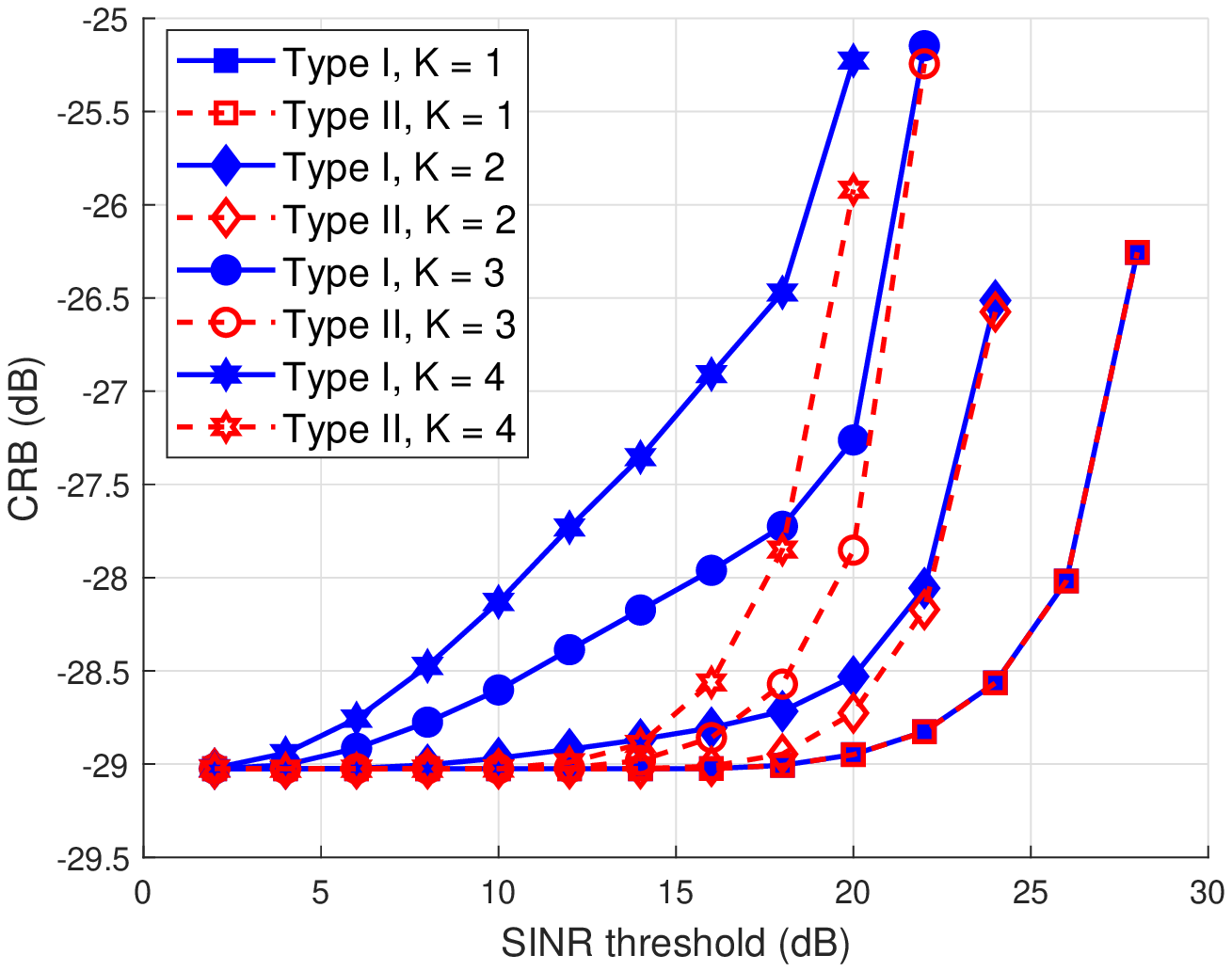}
		 \caption{The CRB for target response matrix estimation by the proposed designs versus the SINR threshold $\Gamma$ with extended target.}
		 \label{fig:CRB_proposed}
	\end{minipage}
	\quad
	\begin{minipage}{0.45\linewidth}
		\centering
		\includegraphics[width=0.9\linewidth]{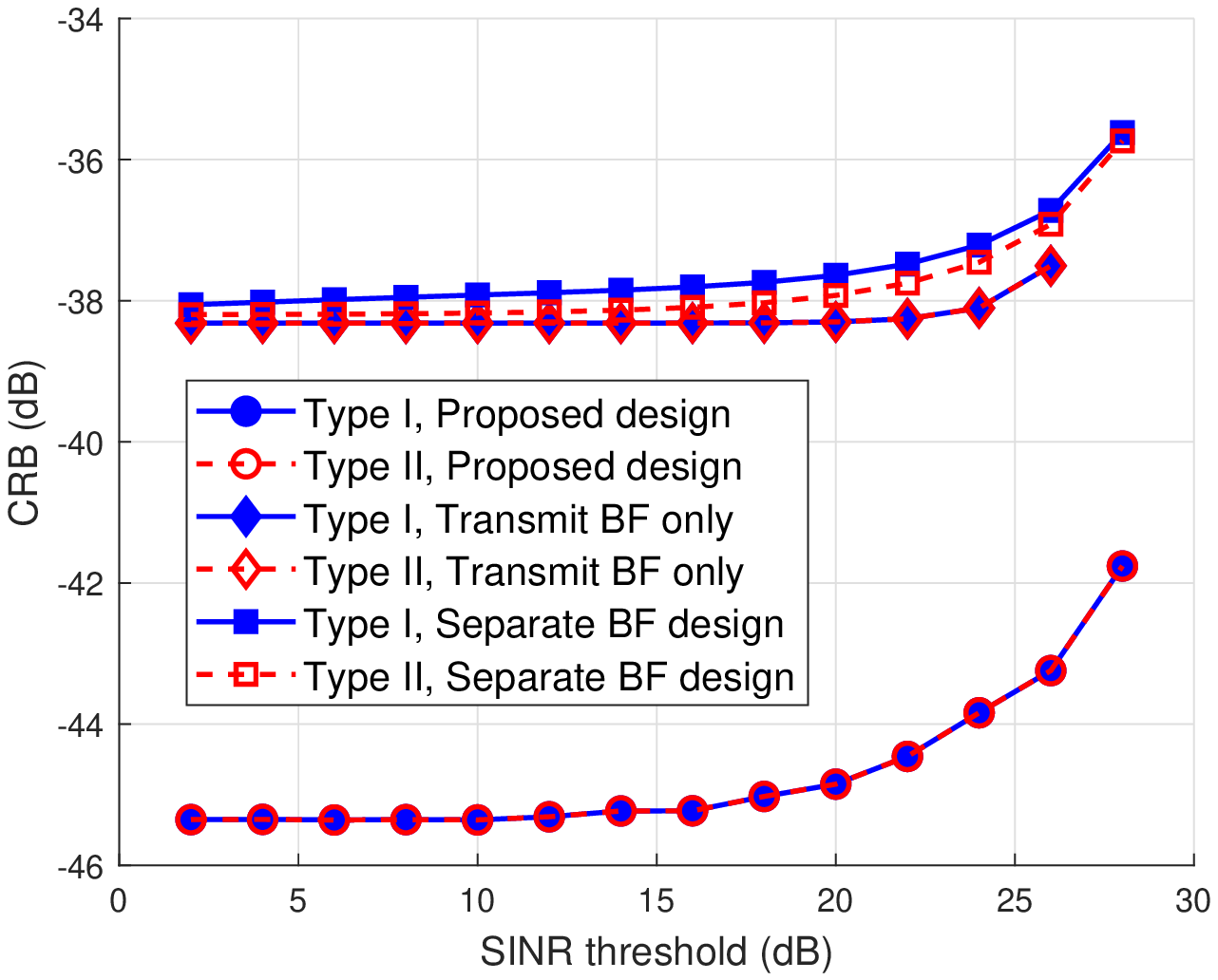}
		 \caption{The CRB for DoA estimation versus the SINR threshold $\Gamma$ with point target, where $K=1$.}
    	\label{fig:CRB_SNR_single_point}
	\end{minipage}
\end{figure}

% \begin{figure}[t]
%     \centering
%     \includegraphics[width=0.38\textwidth]{Fig/CRB_vs_SNR_K_1.eps}
%     \caption{The estimation CRB versus the SINR threshold $\Gamma$ with $K=1$.}
%     \label{fig:CRB_SNR_single}
% \end{figure}
% \begin{figure}[t]
%     \centering
%     \includegraphics[width=0.38\textwidth]{Fig/CRB_vs_SNR_K_3.eps}
%     \caption{The estimation CRB versus the SINR threshold $\Gamma$ with $K=3$.}
%     \label{fig:CRB_SNR_multi}
% \end{figure}
%\subsection{Extended Target Case}
First, we evaluate the ISAC performance with extended target.
Fig.~\ref{fig:CRB_SNR_single} and Fig.~\ref{fig:CRB_SNR_multi} show the estimation CRB versus the SINR threshold $\Gamma$ when $K=1$ and $K=4$, respectively. It is observed in Fig.~\ref{fig:CRB_SNR_single} that with $K=1$, the proposed design (or the transmit BF only) with Type-I CU receivers achieves the same CRB as that with Type-II CU receivers. This can be explained based on Propositions \ref{prop:single_user_Type-I} and \ref{prop:single-user}.  It is also observed that for both types of receivers, the proposed designs outperform other benchmark schemes, which shows the benefit of our proposed designs with jointly optimized communication and sensing beamformers as well as IRS reflective beamformer.
When the SINR threshold $\Gamma$ is sufficiently small, the estimation CRB of the benchmark schemes is close to that of the proposed designs. This is because that in this case, the SINR requirements at the CU receivers are easily satisfied, and thus the beamforming design is sensing oriented, and the CRB for target response matrix estimation only depends on the transmit beamforming.
When the SINR threshold $\Gamma$ is sufficiently large, the transmit BF only design is inferior to the separate BF design or even infeasible, as the communication SINR becomes the performance bottleneck. It is also observed that with moderate values of $\Gamma$, the transmit BF only outperforms the separate BF design. This is due to the fact that the estimation CRB only depends on the transmit beamforming, which is thus more important in this case.

Fig.~\ref{fig:CRB_proposed} shows the estimation CRB achieved by the proposed designs versus the SINR threshold $\Gamma$ under different values of $K$ with extended target. Besides the similar observations as for Figs.~\ref{fig:CRB_SNR_single} and \ref{fig:CRB_SNR_multi}, it is observed that when $K$ becomes large, the performance gap between the two types of CU receivers becomes more significant in the regime of moderate $\Gamma$. Furthermore, when $\Gamma$ is sufficiently high, their performance gap is observed to become marginal. This is because in this case, most power should be allocated to communication signals to ensure the SINR constraints at CU receivers, thus the sensing signal power and the resultant interference become negligible.

% It is also observed that for the case of moderate values of $\Gamma$, the use of Type-II CU receivers leads to significant sensing performance enhancement when $K>1$, and the performance gains increased with number of CUs. This shows the benefit of Type-II CU receivers for multiuser ISAC system.

%  proposed design achieves the lowest CRB for each type of CU receivers
% the same performance for both types of ID receivers is observed
% in Fig. 4. On the other hand, when is sufficiently large, this
% case corresponds to Region 1 in Fig. 3, in which it is optimal
% to allocate all transmit power to information beams to ensure
% that the SINR constraints at ID receivers are all met; as a result, transmit power allocated to energy beams is zero for both
% types of ID receivers, and thus their performances are also identical. At last, for the case of moderate values of which corresponds to Region 2 in Fig. 3, the considerable performance gain
% by Type II over Type I ID receivers is due to the use of one dedicated energy beam. For example, under this particular channel
% setup, as shown in Fig. 4, a 41% average harvested power gain is
% achieved for EH receivers with Type II ID receivers as compared
% to Type I ID receivers when and , thanks to
% the cancellation of (known) energy signals at ID receivers.

%\subsection{Point Target Case}
\begin{figure}[t]
	\centering
	\begin{minipage}{0.45\linewidth}
		\centering
		\includegraphics[width=0.9\linewidth]{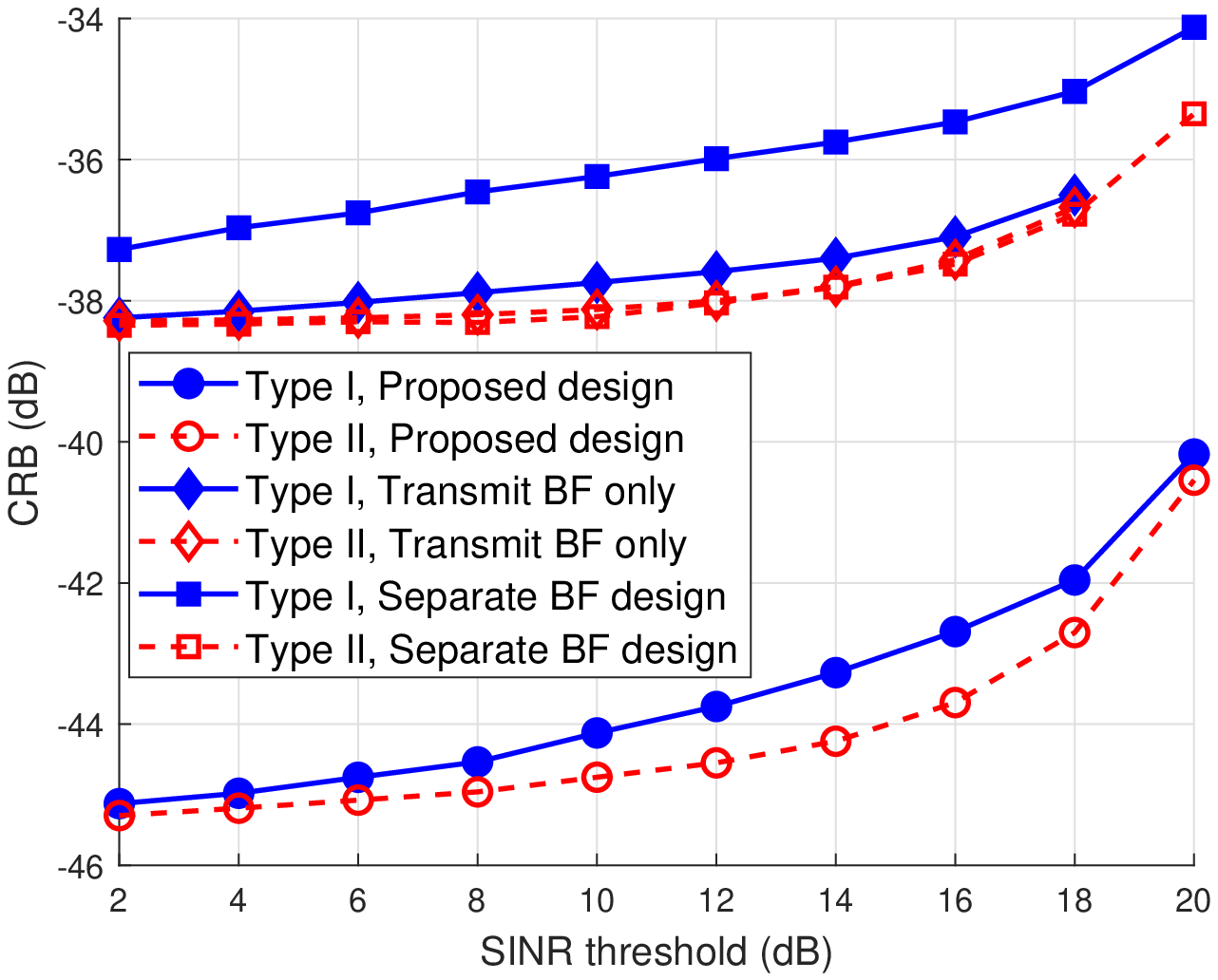}
		\caption{The CRB for DoA estimation versus the SINR threshold $\Gamma$ with point target, where $K=4$.}
    	\label{fig:CRB_SNR_multi_point}
	\end{minipage}
	\quad
	\begin{minipage}{0.45\linewidth}
		\centering
		\includegraphics[width=0.9\linewidth]{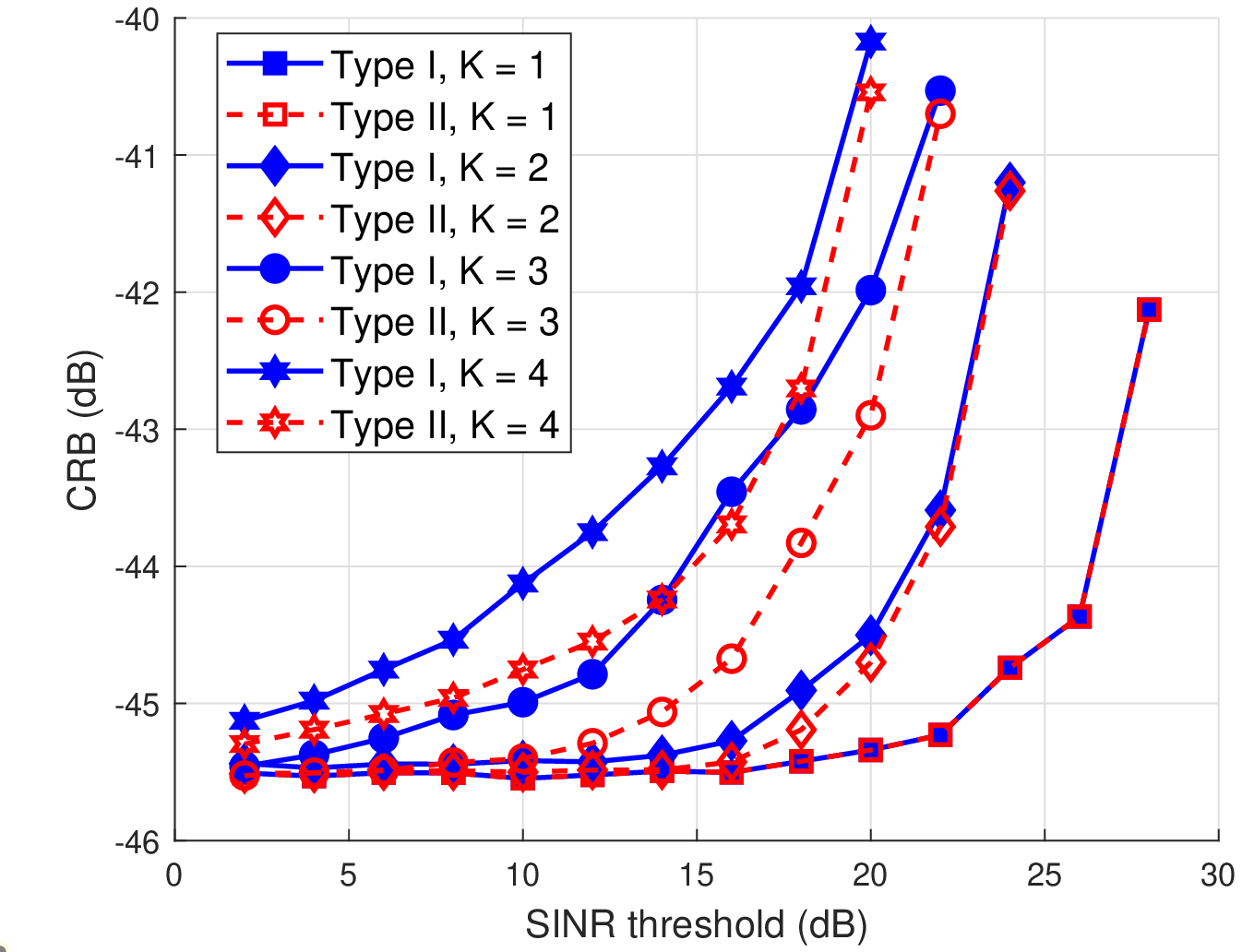}
		 \caption{The CRB for target response matrix estimation by the proposed designs versus the SINR threshold $\Gamma$ with point target.}
		 \label{fig:CRB_proposed_point}
	\end{minipage}
\end{figure}

Next, we evaluate the ISAC performance with point target. Fig.~\ref{fig:CRB_SNR_single_point} and Fig.~\ref{fig:CRB_SNR_multi_point} show the estimation CRB versus the SINR threshold $\Gamma$ when $K=1$ and $K=4$, respectively. It is observed that the proposed designs outperform other benchmark schemes significantly, which shows the benefit of our proposed joint beamforming designs. It is also observed that when the SINR threshold $\Gamma$ is sufficiently large, the transmit BF only design is infeasible.

Fig.~\ref{fig:CRB_proposed_point} shows the estimation CRB achieved by the proposed designs versus the SINR threshold $\Gamma$ under different values of $K$ with point target.  It is also observed that the CRB performance of Type-II CU receivers is identical with that with Type-II CU receivers when $K=1$, and outperforms that with Type-II CU receivers when $K>1$. This is consistent with Proposition~\ref{prop:single_user_Type-I_point} and shows the benefit of Type-II CU receivers when there are more than one CUs. 
Furthermore, it is observed that the performance gap becomes marginal when $\Gamma$ is sufficiently low or high, and the gap becomes significant for the moderate values of $ \Gamma$. This observation can be similarly explained as for Fig.~\ref{fig:CRB_proposed}.

% When $\Gamma$ is sufficiently small, the SINR requirements at CU receivers are easily satisfied, and thus the beamforming design is sensing oriented, then their performance gap becomes marginal.
% When sufficiently $\Gamma$ is sufficiently large, most power should be allocated to information signals to ensure the SINR constraints at CU receivers are all met, thus making the power allocated to dedicated sensing signals is relatively small for both types of receivers, and thus their performance gap also becomes marginal. 

\section{Conclusion}
This paper studied the joint transmit and reflective beamforming design in an IRS-enabled single-target multi-CUs ISAC system, by considering two types of targets, namely extended and point targets. We considered that the BS sends dedicated sensing signals in addition to the communication signals, and accordingly considered two types of CU receivers without and with the sensing signal interference cancellation capability, respectively. Under the different target models and different CU receiver types, we proposed to minimize the estimation CRB by jointly optimizing the transmit and reflective beamforming, subject to the minimum SINR constraints at individual CUs and maximum transmit power constraint at the BS. Numerical results showed that the joint beamforming design can significantly enhance the ISAC performance in terms of minimized CRB as compared to other benchmark schemes without  joint optimization, while the dedicated sensing signals and sensing signal interference cancellation at CU receivers are crucial for performance improvement when the number of CUs is greater than one. 

\appendix
\subsection{Proof of Proposition \ref{prop:SDR}} \label{sec:proof_of_proposition_SDR}
Define $\mathbf W_k^\text{opt,I} = \mathbf w_k^\text{opt,I} (\mathbf w_k^\text{opt,I})^{H} \succeq \mathbf 0, \forall k \in \mathcal{K}$. We prove this proposition by showing that $\{\mathbf W_k^\text{opt,I}\}$ (of rank one) and $\mathbf R_0^\text{opt,I}$ are also optimal for (SDR1.2). 
%Based on Proposition~\ref{prop:SDR}, the optimal solution to problem (SDR1.2) is $\{\mathbf W_k^\star\}$ and $\mathbf R_0^\star$. 
It follows from \eqref{eq:R_new} that $\sum_{k\in\mathcal{K}}\mathbf W_k^\text{opt,I}+\mathbf R_0^\text{opt,I} = \sum_{k\in\mathcal{K}} \mathbf W_k^\star + {\mathbf R}_0^\star$. As a result, the objective value and transmit power obtained by $\{\mathbf W_k^\text{opt,I}\}$ and $\mathbf R_0^\text{opt,I}$ remains the same for (SDR1.2). It can be verified that $\mathrm {tr}(\mathbf H_k  \mathbf W_k^\text{opt,I}) = \mathrm {tr}(\mathbf H_k  \mathbf W_k^\star), \forall k \in \mathcal{K}$.
Therefore,  the SINR constraints in \eqref{eq: SINR_W_I_reformulated} are met.
Furthermore, for any $\mathbf y \in \mathbb{C}^{M \times 1}$, it holds that
\begin{equation}
\mathbf y^{H} ({\mathbf W}_k^\star - \mathbf W_k^\text{opt,I})\mathbf y = \mathbf y^{H} {\mathbf W}_k^\star  \mathbf y - |\mathbf y^H {\mathbf W}_k^\star  \mathbf h_k|^2(\mathbf h_k^H {\mathbf W}_k^\star \mathbf h_k)^{-1}.
\end{equation}
According to the Cauchy-Schwarz inequality, we have
$(\mathbf y^H {\mathbf W}_k^\star \mathbf y) (\mathbf h_k^{H} {\mathbf W}_k^\star \mathbf h_k)   \geq |\mathbf y^H {\mathbf W}_k^\star \mathbf h_k|^2$,
and thus it follows that $\mathbf y^H ({\mathbf W}_k^\star - \mathbf W_k^\text{opt,I})\mathbf y\geq 0$. Accordingly, we have  ${\mathbf W}_k^\star - \mathbf W_k^\text{opt,I} \succeq \mathbf 0,\forall k \in \mathcal{K}$. It follows from \eqref{eq:R_new} that $\mathbf R_0^\text{opt,I} \succeq \mathbf 0$. Hence, $\{\mathbf W_k^\text{opt,I} \}$ and $\mathbf R_0^\text{opt,I}$ are optimal for (SDR1.2).  Proposition~\ref{prop:SDR} is thus proved.

\subsection{Proof of Proposition \ref{proposition_zero}} \label{sec:proof_of_proposition_zero}
Let $\{\hat{\mathbf W}_k^\star\}$ and $\hat{\mathbf R}_0^\star$ denote the optimal solution of (SDR1.2), where $\hat{\mathbf R}_0^\star$ is non-zero in general. Then, we choose arbitrary $i^\star \in\mathcal{K}$, and reconstruct an alternative solution as $\mathbf R_0^\star = \mathbf 0$ and
\begin{equation}
\mathbf W_k^\star = \begin{cases} \hat{\mathbf W}_k^\star+\hat{\mathbf R}_0^\star & k=i^\star,\\
\hat{\mathbf W}_k^\star & k\neq i^\star.
\end{cases} 
\end{equation}
It is easy to show that $\sum_{k\in\mathcal{K}}\mathbf  W_k^\star+\mathbf  R_0^\star=\sum_{k\in\mathcal{K}}\hat{\mathbf  W}_k^\star+\hat{\mathbf R}_0^\star$
 and
 % $\mathrm{tr}(\mathbf  H_k\hat{\mathbf  W}_k) \ge \mathrm{tr}(\mathbf  H_k\mathbf  W_k^\star), \forall k \in \mathcal{K}$.
\begin{equation}
\mathrm{tr}(\mathbf  H_k\mathbf  W_k^\star) = \begin{cases}  \mathrm{tr}(\mathbf  H_k\hat{\mathbf  W}_k^\star) + \mathrm{tr}(\mathbf  H_k\hat{\mathbf R}_0^\star)  \ge \mathrm{tr}(\mathbf  H_k\hat{\mathbf  W}_k^\star) & k=i^\star,\\
\mathrm{tr}(\mathbf  H_k\hat{\mathbf  W}_k^\star) & k\neq i^\star.
\end{cases} 
\end{equation}
Therefore, the alternative solution $\{\mathbf W_k^\star\}$ and  $\mathbf R_0^\star$ satisfy all the constraints and achieve the same objective value for (SDR1.2) as that by $\{\hat{\mathbf W}_k^\star\}$ and $\hat{\mathbf R}_0^\star$. As a result, the reconstructed solotion $\{\mathbf W_k^\star\}$ and $\mathbf R_0^\star=\mathbf 0$ are optimal for (SDR1.2).  Proposition~\ref{proposition_zero} is thus proved.

\subsection{Proof of Proposition \ref{prop:SDR_2}} \label{sec:proof_of_proposition_1_2}
Define $\mathbf W_k^\text{opt,II} = \mathbf w_k^\text{opt,II} ( \mathbf w_k^\text{opt,II})^{H}\succeq \mathbf 0, \forall k \in \mathcal{K}$. We prove this proposition by showing that $\{\mathbf W_k^\text{opt,II}\}$ (of rank one) and $\mathbf R_0^\text{opt,II}$ are also optimal for (SDR2.1). 
Similar to (SDR1.2),  under the reconstructed solution, the objective value remains the same and the constraints in \eqref{eq: power_W} and~\eqref{eq:W_semi} are met. We have $\mathrm {tr}(\mathbf H_k \mathbf W_k^\text{opt,II}  )=\mathrm {tr}(\mathbf H_k \mathbf W_k^{\star\star} )$ and $\mathbf W_k^{\star\star}- \mathbf W_k^\text{opt,II}  \succeq \mathbf 0, k \in \mathcal{K}$,
 such that
\begin{equation}
\frac{1}{\Gamma_k}\mathrm{tr}(\mathbf  H_k\mathbf W_k^\text{opt,II}  )-\sum_{i\in\mathcal{K},i\neq k}\mathrm{tr}(\mathbf  H_k\mathbf W_i^\text{opt,II}  )\ge\frac{1}{\Gamma_k}\mathrm{tr}(\mathbf  H_k\mathbf W_k^{\star\star})-\sum_{i\in\mathcal{K},i\neq k}\mathrm{tr}(\mathbf  H_k\mathbf W_i^{\star\star})
\ge\sigma_k^2.
\end{equation}
Then $\{\mathbf W_k^\text{opt,II}\}$ and $\mathbf R_0^\text{opt,II}$ also satisfy the SINR constraints in \eqref{eq: SINR_W_II}, and thus are optimal for (SDR2.1).  Proposition \ref{prop:SDR_2} is thus proved.

\ifCLASSOPTIONcaptionsoff
  \newpage
\fi

\bibliographystyle{IEEEtran}
\bibliography{IEEEabrv,mybibfile}

\end{document}